\documentclass[journal]{IEEEtran}

\ifCLASSINFOpdf
  \usepackage[pdftex]{graphicx}
  \graphicspath{{./figures/}}
  \DeclareGraphicsExtensions{.pdf,.jpeg,.png,.eps}
\else
\fi

\usepackage{color}
\usepackage{pifont}
\usepackage{multirow}
\usepackage{xspace}
\usepackage{hyperref}
\usepackage{fontawesome5}
\usepackage{academicons} 
\usepackage{array} 
\usepackage{makecell} 
\usepackage{rotating}
\usepackage{balance}
\newcommand{\cnc}{CNC\xspace}
\newcommand{\cmark}{\tiny\ding{108}}
\newcommand{\xmark}{\phantom{\cmark}}

\usepackage{amsmath}

\newcommand{\Artery}{\href{https://github.com/riebl/artery}{Artery}\xspace} 
\newcommand{\CloudSim}{\href{https://github.com/Cloudslab/cloudsim}{CloudSim}\xspace} 
\newcommand{\CloudSimPlus}{\href{https://github.com/cloudsimplus/cloudsimplus}{CloudSim Plus}\xspace} 
\newcommand{\CloudSimExpress}{\href{https://github.com/Cloudslab/cloudsim-express}{CloudSim Express}\xspace} 
\newcommand{\CloudSimPlusAutomation}{\href{https://github.com/manoelcampos/cloudsimplus-automation}{CloudSim Plus Automation}\xspace} 
\newcommand{\CloudSimSDN}{\href{https://github.com/Cloudslab/cloudsimsdn}{CloudSimSDN}\xspace} 
\newcommand{\CloudSimPy}{\href{https://github.com/FengcunLi/CloudSimPy/blob/master/README.en.md}{CloudSimPy}\xspace} 
\newcommand{\CFN}{\href{https://github.com/spirosmastorakis/CFN/}{CFN}\xspace} 
\newcommand{\Cooja}{\href{https://github.com/contiki-ng/cooja}{Contiki Cooja}\xspace} 
\newcommand{\CORE}{\href{https://github.com/coreemu/core}{CORE}\xspace} 
\newcommand{\DFaaS}{\href{https://github.com/UNIMIBInside/dfaas}{DFaaS}\xspace} 
\newcommand{\EasiEI}{\href{https://gitlab.com/Mirrola/ns-3-dev}{EasiEI}\xspace} 
\newcommand{\ECSNeT}{\href{https://github.com/sedgecloud/ECSNeTpp}{ECSNeT++}\xspace} 
\newcommand{\EdgeCloudSim}{\href{https://github.com/CagataySonmez/EdgeCloudSim}{EdgeCloudSim}\xspace} 
\newcommand{\SimEdgeIntel}{\href{https://github.com/XiaofeiTJU/SimEdgeIntel}{SimEdgeIntel}\xspace} 
\newcommand{\EmuFog}{\href{https://github.com/emufog/emufog}{EmuFog}\xspace} 
\newcommand{\faassim}{\href{https://github.com/edgerun/faas-sim}{faas-sim}\xspace} 
\newcommand{\Fogify}{\href{https://ucy-linc-lab.github.io/fogify/}{Fogify}\xspace} 
\newcommand{\iFogSim}{\href{https://github.com/Cloudslab/iFogSim}{iFogSim2}\xspace} 
\newcommand{\IoTSimEdge}{\href{https://github.com/DNJha/IoTSim-Edge}{IoTSim-Edge}\xspace}
\newcommand{\IoTSimOsmosis}{\href{https://github.com/kalwasel/IoTSim-Osmosis}{IoTSim-Osmosis}\xspace} 
\newcommand{\LEAF}{\href{https://github.com/dos-group/leaf}{LEAF}\xspace} 
\newcommand{\MaxiNet}{\href{https://github.com/MaxiNet/MaxiNet}{MaxiNet}\xspace} 
\newcommand{\MobFogSim}{\href{https://github.com/diogomg/MobFogSim}{MobFogSim}\xspace} 
\newcommand{\MiniNDN}{\href{https://github.com/named-data/mini-ndn}{Mini-NDN}\xspace} 

\newcommand{\MinNet}{\href{http://mininet.org/}{MiniNet (WiFi)} \xspace} 
\newcommand{\ndnSIM}{\href{https://ndnsim.net/}{ndnSIM}\xspace} 
\newcommand{\NSthree}{\href{https://www.nsnam.org}{NS-3}\xspace} 
\newcommand{\OMNeT}{\href{https://github.com/omnetpp/omnetpp}{OMNeT++}\xspace} 
\newcommand{\PacketTracer}{\href{https://www.netacad.com/courses/packet-tracer/introduction-packet-tracer}{Cisco Packet Tracer}\xspace} 
\newcommand{\PeerSim}{\href{http://peersim.sourceforge.net/}{PeerSim}\xspace} 
\newcommand{\PureEdgeSim}{\href{https://github.com/CharafeddineMechalikh/PureEdgeSim}{PureEdgeSim}\xspace} 
\newcommand{\RECAPDES}{\href{https://bitbucket.org/RECAP-DES/recap-des/src/master/}{RECAP-DES}\xspace} 

\newcommand{\SatEdgeSim}{\href{https://github.com/wjy491156866/SatEdgeSim}{SatEdgeSim}\xspace} 
\newcommand{\Shadow}{\href{https://github.com/shadow/shadow}{Shadow}\xspace} 
\newcommand{\Phantom}{\href{https://github.com/shadow/shadow}{Phantom}\xspace} 
\newcommand{\SimFaaS}{\href{https://github.com/pacslab/simfaas}{SimFaaS}\xspace} 
 
\newcommand{\SimGrid}{\href{https://github.com/simgrid/simgrid}{SimGrid}\xspace} 

\newcommand{\SimufifthG}{\href{https://github.com/Unipisa/Simu5G}{Simu5G}\xspace}

\newcommand{\StarryNet}{\href{https://github.com/SpaceNetLab/StarryNet}{StarryNet}\xspace}
\newcommand{\StepONE}{\href{https://github.com/jaks6/step-one}{Step-ONE}\xspace}
\newcommand{\TheONE}{\href{https://akeranen.github.io/the-one/}{The ONE}\xspace} 
\newcommand{\VirtFogSim}{\href{https://github.com/mscarpiniti/VirtFogSim}{VirtFogSim}\xspace} 
\newcommand{\YAFS}{\href{https://github.com/acsicuib/YAFS}{YAFS}\xspace} 

\newcommand{\Java}{\includegraphics[height=2ex]{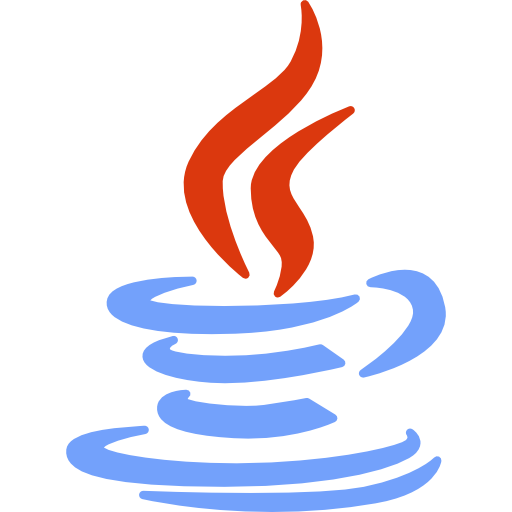}\xspace}
\newcommand{\Python}{\includegraphics[height=2ex]{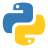}\xspace}
\newcommand{\Matlab}{\includegraphics[height=2ex]{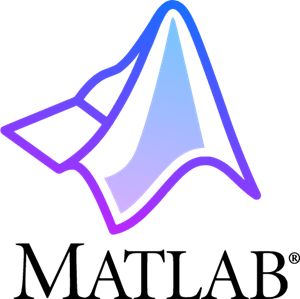}\xspace}
\newcommand{\YAML}{\includegraphics[height=2ex]{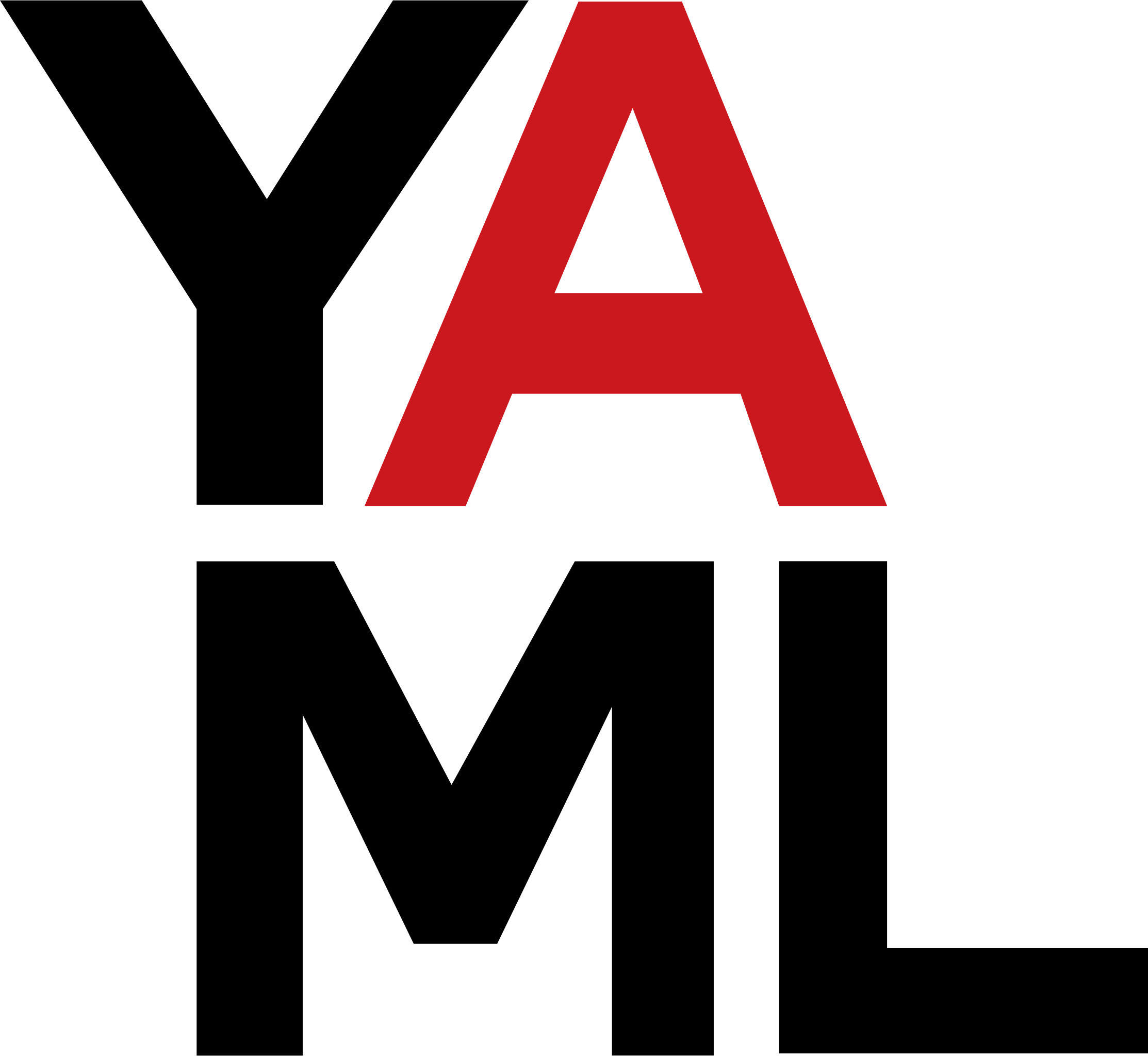}\xspace}
\newcommand{\CPP}{\includegraphics[height=2ex]{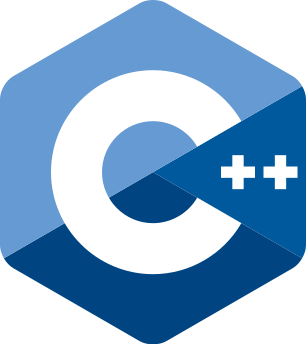}\xspace}
\newcommand{\Shell}{\includegraphics[height=2ex]{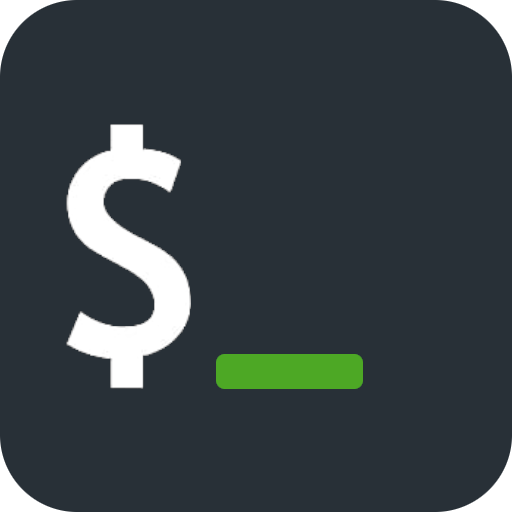}\xspace}
\newcommand{\Kotlin}{\includegraphics[height=2ex]{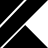}\xspace}
\newcommand{\Rust}{\includegraphics[height=2ex]{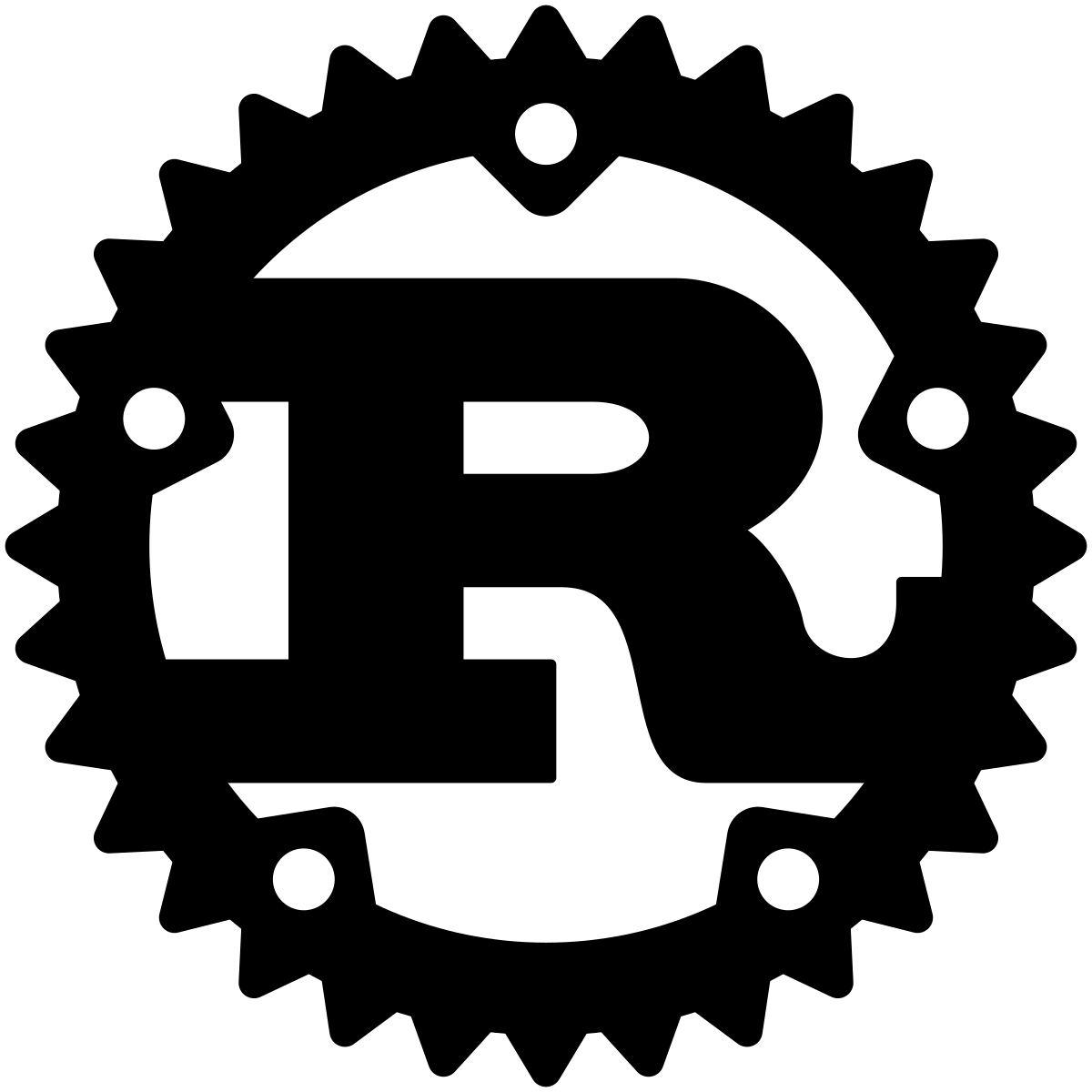}\xspace}

\newcommand{\AwesomeEdgeComputing}{\url{https://github.com/qijianpeng/awesome-edge-computing}}

\begin{document}
\title{A Survey on Open-Source Edge Computing Simulators and Emulators: The Computing and Networking Convergence Perspective}
%
%
%

\author{Jianpeng~Qi, 
        Chao~Liu,
        Xiao~Zhang, %
        Lei~Wang, 
        Rui~Wang, 
        Junyu Dong, 
        and~Yanwei~Yu %
\thanks{Jianpeng~Qi, Chao~Liu, Xiao~Zhang, Junyu Dong, and Yanwei~Yu are with the Faculty of Information Science and Engineering, Ocean University of China, Qingdao, China. }
\thanks{Rui~Wang is with the School of Computer and Communication Engineering, University of Science and Technology Beijing, Beijing, China.}
\thanks{Lei~Wang is with the State Grid Intelligence Technology Co., Ltd., Jinan, China.}
\thanks{Correspondence: Yanwei~Yu.}
\thanks{Manuscript received Dec 19, 2023; revised xx xx, 20xx.}}

%
%

\markboth{Journal of \LaTeX\ Class Files,~Vol.~14, No.~8, August~2015}%
{Shell \MakeLowercase{\textit{et al.}}: Bare Demo of IEEEtran.cls for IEEE Journals}
%



\maketitle

\begin{abstract}
Edge computing, with its low latency, dynamic scalability, and location awareness, along with the convergence of computing and communication paradigms, has been successfully applied in critical domains such as industrial IoT, smart healthcare, smart homes, and public safety. This paper provides a comprehensive survey of open-source edge computing simulators and emulators, presented in our GitHub repository (\AwesomeEdgeComputing), emphasizing the convergence of computing and networking paradigms. By examining more than 40 tools, including CloudSim, NS-3, and others, we identify the strengths and limitations in simulating and emulating edge  environments.
This survey classifies these tools into three categories: packet-level, application-level, and emulators. Furthermore, we evaluate them across five dimensions, ranging from resource representation to resource utilization. The survey highlights the integration of different computing paradigms, packet processing capabilities, support for edge environments, user-defined metric interfaces, and scenario visualization. The findings aim to guide researchers in selecting appropriate tools for developing and validating advanced computing and networking technologies.
\end{abstract}


%
\section{Introduction}
\label{sect_intro}
\IEEEPARstart{E}{dge} computing, with features such as low latency, dynamics,
mobility, and location awareness, has been successfully applied in critical
domains, such as industrial Internet of Things (IoT), smart healthcare, smart
homes, and public safety, yielding significant economic benefits and influencing
various aspects of daily life \cite{RN935,RN940}. As emerging network
technologies continue to evolve, the word \textit{``future network''}  can be
seen almost everywhere, such as information-centric networking \cite{RN900},
in-networking computing \cite{RN1315}, compute first networking \cite{RN25,10172050},
computing power network \cite{RN93}, and smart identifier network
\cite{7474343}. In such networks, a myriad of computing nodes interconnect,
forming a network with substantial computing capabilities, thereby optimizing
the utilization of both communication and computing resources. As a result, the
location of computation has shifted from occurring solely at the network edge to
a continuous integration of computing resources between the cloud and the edge
\cite{7488250}.

We can now draw a clear picture that achieving \textit{computing and network
	convergence} (\cnc~for short) becomes a trending \cite{10304187,RN8}.  %
\cnc as the integrated design and operation of communication and computation processes within tasks, where resource allocation is managed jointly to enhance overall system performance under a unified objective, such as minimizing energy consumption, reducing latency, or maximizing computation accuracy \cite{RN1604, RN93}. For instance, in scenarios such as smart venues, remote surgeries, and
industrial IoT, terminal devices on-site can perform lightweight pre-processing
on various types of data and transmit the intermediate results into the network.
By adopting the concept of \cnc, intelligent tasks that are resource-intensive,
such as data aggregation and analysis, image enhancement, can be executed by
computing-capable nodes (e.g., smart gateways, intelligent accelerator cards)
along the path to user terminals. This approach achieves ultra-low latency for
analysis and results delivery \cite{RN8,9862944}.

Under this trend, numerous open-source organizations and projects have emerged,
establishing robust systems and software stacks to maximize the value of \cnc.
For instance, LF Edge \cite{RN1339} aims to create an open and inter-operable
edge computing framework by providing a unified platform and standards. Eclipse
IoT \cite{RN1340}, focusing on the IoT domain, strives to offer universal tools
and frameworks for connecting, managing, and monitoring IoT devices and
applications. OpenStack StarlingX \cite{RN1341}, emphasizing rapid response and
recovery, endeavors to build a secure, reliable, and low-latency edge cloud
environment for critical tasks. CNCF KubeEdge \cite{RN1342}, extending
Kubernetes container orchestration to edge nodes, aims to run containerized
applications and services on edge devices, ensuring reliable management and
monitoring to accelerate cloud-edge collaboration. We also maintain a more
comprehensive list of open-source projects in \cite{awesome}.

However, from the perspective of academia, the types and the scale of networks
are diverse, causing the cost of validating ideas in large-scale real-world
scenarios become high. Therefore, simulation tools become enabling technology
not only for researching but also for teaching (such as \PacketTracer).
Recently, many novel simulators have been implemented with \cnc required
features, such as ndnSIM \cite{ndnsim}, iFogSim \cite{RN1331}, and EasiEI
\cite{easiei}. This paper aims to study the open-source simulators that exist in or relate to
\cnc, analyze the characteristics of these platforms, identify common features,
and provide insight into promising directions for future development.

\subsection{Challenges When Simulating \cnc}
\begin{figure*}
	\centering
	\includegraphics[width=0.9\textwidth]{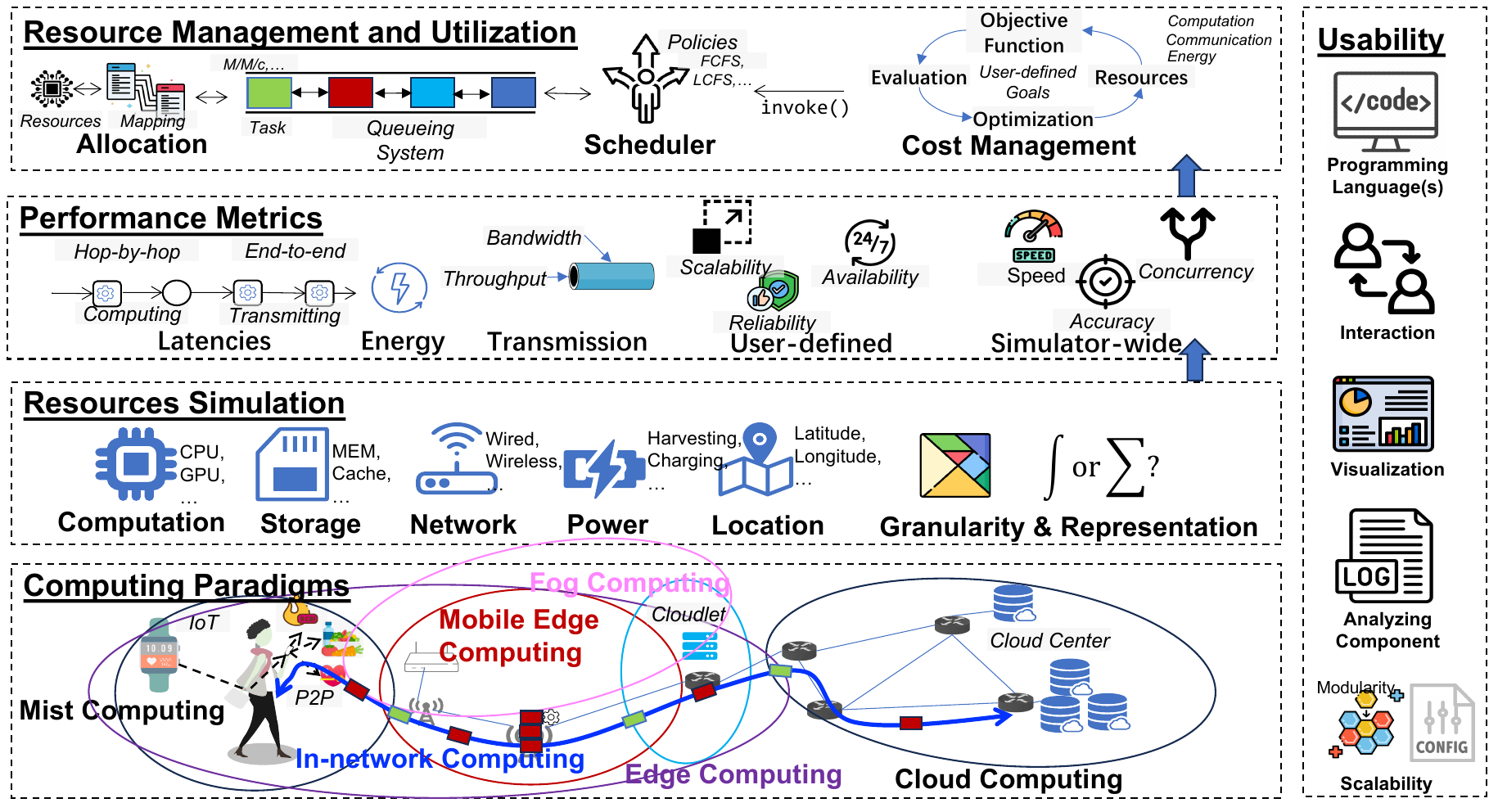}
	\caption{Categories Classified in This Survey}
	\label{fig:categories}
\end{figure*}

Simulating \cnc scenarios involves many challenges, such as different computing paradigms, heterogeneous resources, and various service metrics. After carefully reviewing the characteristics of multiple reviews and simulators, we have grouped the features of open-source simulators into five main dimensions (see \figurename~\ref{fig:categories}): \textit{Computing Paradigms, Resource Simulation, Performance Metrics, Resource Management and Utilization, and Usability}.

\subsubsection{\textbf{Computing Paradigms}}
The term Computing Paradigms refers to distinct models or approaches to processing and managing data and resources in distributed environments. These paradigms include, but are not limited to, cloud computing, fog computing, edge computing, and peer-to-peer (P2P) computing. Each paradigm has its unique characteristics in terms of infrastructure, access technologies, and resource management strategies. Integrating these diverse computing paradigms into a unified environment is challenging due to the significant variations in how resources are handled and accessed. These differences complicate interoperability and hinder seamless collaboration among paradigms, as resources managed within one paradigm are not easily shared with others, making it difficult to create a cohesive resource pool \cite{10138337, 10091816}. Addressing this heterogeneity is essential for improving resource utilization across edge environments.

Traditional simulators, like CloudSim \cite{cloudsim}, were originally designed for cloud computing environments and rely on fixed abstractions such as centralized computing models, which are not well-suited for the dynamic, decentralized, and hierarchical needs of newer paradigms like fog and edge computing. Over time, CloudSim has evolved with extensions such as iFogSim \cite{ifogsim2} and EdgeCloudSim \cite{edgecloudsim} to better support the complexity and diversity of edge computing, yet challenges remain in accommodating the full range of paradigms. For example, in the Internet of Vehicles (IoV) scenarios, vehicles may need to dynamically switch between peer-to-peer communication for localized data exchange and centralized cloud computing for broader data aggregation and analysis. The challenge, therefore, is to develop a flexible simulator that not only integrates these diverse paradigms but also adapts to their unique requirements, providing accurate and reliable results across various scenarios.

Meanwhile, the limitation of selecting a single paradigm in simulator design may  lead to constraints in future adaptability, as computing architectures are continuously evolving. Traditional cloud computing techniques, for example, may not be entirely applicable to edge computing due to emerging considerations such as geo-distribution and privacy issues \cite{7488250}. Simulators developed with a centralized approach may not seamlessly transition to the current demands of network-wide computing and resource transmission. Furthermore, in modern network architectures, every intermediate node can potentially serve as a computing node, either at the data packet or application level. These nodes can also be tasked with various forms of offloading \cite{RN25, 7474343, RN1315}.

Therefore, this survey aims to address a crucial question: \textit{What types of computing paradigms are supported by different simulators?} 
By understanding the range of paradigms that each simulator supports, researchers can make informed decisions when selecting a simulation platform that best fits their specific needs. This insight is not only valuable for those seeking dedicated platforms to support their current research but also for those looking to enhance their simulators by integrating diverse computing paradigms, thus enabling more comprehensive and flexible simulations in rapidly evolving computational environments.

\subsubsection{\textbf{Resources Simulation}}
Resource Simulation refers to the process of modeling and emulating the behavior, performance, and interactions of various hardware and software resources within a computing environment. In \cnc environments, this involves replicating the operational characteristics of diverse devices and their internal components, such as CPUs, GPUs, memory, network interfaces, and energy sources. Effective simulation of resources in these environments requires careful attention to the heterogeneous and distributed nature of edge computing. Unlike cloud computing environments where resources are centralized, virtualized, and well-defined, edge environments feature a wide array of devices that are geographically dispersed and varied in their capabilities. These devices can include intelligent terminals, desktops, laptops, gateways, WiFi access points (APs), and more, each with different computational and networking capacities \cite{resmgmtsurvey, RN93, 7488250}.

When simulating these diverse devices, one approach is to run the instructions, capturing the nuances of the device’s configuration and state. This allows for detailed modeling of components such as CPUs, GPUs, memory, and network interfaces, and it accurately reflects real-world conditions, like gem5 \cite{gem5}. Alternatively, more approximate methods, such as counting the number of devices or calculating CPU cycles, can be used for broader, less detailed simulations \cite{cloudsim}.

For instance, simulating a real-world edge computing scenario not only involves accounting for the variability in device types and capabilities but also the dynamic nature of resource allocation and environmental factors like temperature and signal interference \cite{ASLANPOUR2020100273}. While current simulators may use simplified models to represent these resources, they often lack the granularity needed to accurately capture such complexities, potentially leading to less reliable results. Using a coarse granularity may compromise the validity of simulation outcomes, whereas a finer granularity can lead to increased computational costs and complexity \cite{s23073492}.

Thus, balancing detail and computational efficiency is crucial when developing simulators for edge environments, ensuring that both device-level diversity and internal component intricacies are adequately represented.

Another challenge in the realm of \cnc resource simulation is the rapid emergence and evolution of various types of chips or cards, underscoring the necessity for a flexible and dynamic representation of resources. This dynamic nature of technology demands a representation that is not only accurate but also adaptable to the continuous introduction of new hardware and changes in existing technologies.

Therefore, in the context of resource simulation, the question remains:  \textit{What level of detail or granularity is suitable for representing resources? What types of resources does the simulator support?} These questions are not merely technical but also touch on the broader aspects of standardization, collaboration, and future-proofing in the rapidly evolving field of computing technology.

\subsubsection{\textbf{Performance Metrics}}
Performance Metrics refer to quantifiable measures used to assess the efficiency, effectiveness, and overall performance of systems, processes, or applications within a computing environment. In the context of \cnc, these metrics are essential for evaluating how well different scheduling strategies and resource management techniques perform under various conditions. They play a pivotal role in ensuring that systems meet the requirements of real-time and mission-critical applications, where optimal performance is a necessity \cite{ASLANPOUR2020100273, RN2684}. These metrics provide insights into how effectively a simulator can model diverse scenarios and optimize resource usage. Key performance indicators such as end-to-end latency, throughput, energy consumption, and reliability are fundamental to this evaluation, as they reflect the system’s ability to deliver high performance while maintaining stability and resource efficiency.

End-to-end latency measures the time taken for data to travel from its source to its destination. This metric is particularly important for applications that require real-time responses, such as remote surgeries or autonomous driving, where even slight delays can lead to significant consequences. By understanding the latency, researchers can gauge how quickly a system can deliver data and react to changes in the environment.

Throughput, often measured in terms of bandwidth, refers to the volume of data transmitted over a network that is successfully delivered to its destination. It is a critical metric for applications requiring high data transfer rates, such as video streaming or data-intensive computations. High throughput indicates that the network or system can handle large amounts of data efficiently without significant delays or loss.

Energy consumption, typically measured in watts, is another vital metric, especially for edge computing environments where power resources are limited. It represents the amount of power drawn by computing nodes and networking devices to perform tasks. This metric is essential for designing energy-efficient systems, particularly in scenarios where devices are battery-operated or where energy costs need to be minimized. A comparable cost metric for energy consumption is crucial for fair comparisons across different systems and strategies~\cite{10.1145/3626111.3628186}.

Furthermore, advanced users may define custom metrics that are composed of several basic observed scalars \cite{RN1366}. For instance, scalability is a metric that evaluates a system’s ability to adapt to increasing loads without performance degradation. It is crucial for scenarios where the number of users or devices may fluctuate significantly. Reliability \cite{9488707} and availability \cite{5348800} are metrics that determine the likelihood of a system providing continuous service without interruptions. These are fundamental for applications where downtime is unacceptable, such as in financial transactions or emergency response systems. A related metric, response delay, measures how quickly a system can react to requests, which is crucial for maintaining smooth and efficient operations in dynamic environments \cite{10.1145/3487552.3487815}.

In addition to the above, the simulator itself must have performance metrics to evaluate its effectiveness in large-scale network environments. Metrics such as RAM and CPU usage of the host machine during simulation runs are important to ensure that the simulator can handle complex scenarios without becoming a bottleneck.

Clearly, the performance metrics used in \cnc are diverse, ranging from single-variable metrics like bandwidth to multi-variable ones like reliability. For more performances, one can refer to  \cite{ASLANPOUR2020100273}. They can be either predefined or customized by users to suit specific research needs. Therefore, a question that readers might interested in is \textit{what metrics does the simulator support?} The answer
especially important for researchers who want to check the performances of their
ideas at a lower development cost.

\subsubsection{\textbf{Resource Management and Utilization}}
Effective resource management in \cnc environments relies on the integration of resource allocation, task scheduling, and cost management to achieve optimization for specific cost objectives, such as minimizing energy consumption, reducing latency, or balancing computational overhead \cite{10304187}.  The primary challenge lies in integrating these diverse resources within a single framework to support various scenarios, ranging from low-latency edge processing to high-throughput cloud computing.

Resource allocation involves distributing resources at varying levels of granularity—from entire nodes or clusters to individual CPU cycles, memory units, or bandwidth segments. Because resource demands are highly dynamic and sensitive to network conditions, more advanced strategies, such as dynamic load balancing and resource pooling, are needed to ensure effective resource utilization \cite{10304187}. For example, in an edge-cloud environment, computing resources like CPU cycles and memory may need to be dynamically allocated based on real-time network bandwidth availability. Task scheduling further complements resource allocation by determining the order and optimal placement of tasks across the network. Thus, simulators must support flexible and fine-granularity resource allocation models and a wide range of scheduling policies, such as priority-based scheduling for time-sensitive tasks, real-time scheduling for timely execution, and user-defined custom policies. 

Cost management in \cnc involves reducing the expenses related to using computing and networking resources. In a distributed environment, managing costs becomes more complex because it requires balancing different types of resources. For example, to save energy, some tasks might be migrated to an edge server that has better power efficiency. Since users have different goals and requirements, creating a system that meets all criteria and optimizes costs can be quite challenging \cite{10.1145/3589639}. Therefore, simulators should offer flexible cost models that allow users to set their own criteria and assess the trade-offs between various resource management strategies.

Thus, it is essential to consider the following question when evaluating \cnc simulators: \textit{What resource management and utilization models or policies does the simulator support, and do they offer flexible or user-defined interfaces?} This question is critical because the answer directly impacts the ability of researchers and developers to model complex \cnc scenarios, test various optimization strategies, and ultimately advance the field of edge computing and networking convergence.

\subsubsection{\textbf{Usability}}
The usability of simulation tools is crucial for their adoption in both academic and industrial applications, significantly impacting the ease of setting up, running, and analyzing simulations. Key aspects of usability include visualization capabilities, metrics extraction and analysis, logging, scenario configuration flexibility, and programming language support. For instance, tools like \OMNeT \cite{omnetpp} offer advanced visualization features that help users intuitively understand complex network dynamics, while simulators like \CloudSimExpress \cite{cloudsimexpress} provide user-friendly scripting interfaces to facilitate rapid scenario configuration and modification. These features are essential for researchers to efficiently explore and validate their hypotheses.

Effective metrics extraction and logging are vital for a simulator’s usability, allowing researchers to deeply analyze performance and understand system behavior. Simulators such as \NSthree \cite{nsnam} and its variants support comprehensive metric customization and logging, which are crucial for tasks like debugging, optimization, and performance evaluation. However, achieving a balance between comprehensive data collection and the resulting computational overhead presents a significant challenge. For example, extensive logging can lead to high storage and processing costs, requiring careful design of configurable logging options to manage these trade-offs.

The flexibility of integrating new components or adapting existing ones is another critical usability factor. Simulators that support widely-used programming languages and offer modular, extensible architectures are better suited for evolving research needs. For example, simulators like \NSthree, allow users to extend functionalities easily, which is essential for incorporating novel research areas such as AI-driven resource management (e.g., ns3-gym, a toolkit of using reinforcement learning to solve networking problem \cite{ns3gym}). However, providing this level of flexibility often involves complex engineering efforts to maintain compatibility and stability across diverse modules.

In this paper, we provide a comprehensive survey of the usability features of popular open-source simulators in edge computing and network convergence. We highlight key aspects such as visualization, scenario definition, logging, and configuration flexibility.
Additionally, this survey includes essential information such as the activities, citations, and license types of each simulator, helping researchers make informed decisions when selecting a tool. By offering detailed insights into both the capabilities and limitations of these simulators, we aim to guide researchers in choosing the most suitable platform for their specific needs and to encourage further development in enhancing usability without compromising efficiency.

\subsection{Selection Criteria \& Literature Review Process}
In this study, we systematically track open-source computing and networking simulators/emulators on GitHub, beginning in 2019 \cite{awesome}. Our repository includes a broader array of tools compared to existing surveys, as many of these simulators were identified through research articles and recommendations from GitHub during the development of our EasiEI simulator \cite{easiei}. To date, our repository features over 80 tools, a number that continues to grow as we actively monitor developments in this field.

Given that edge computing and computing and network convergence (CNC) are relatively recent concepts \cite{9108989}, we apply specific criteria to refine the selection of simulators and emulators for this survey. We exclude tools that have not seen code updates for more than six years, focusing instead on those that support general computing paradigms rather than specialized applications, such as UAVs or data caching.

Additionally, we conduct a comprehensive review of the past five years of research using Web of Science and Ei Compendex, employing the query \texttt{"edge computing" AND (simulator OR emulator)} and search within the title/abstract. This search yields a total of 192 results, which are further refined to 50 papers after an initial abstract review. After removing duplicates across the two platforms, we retain 41 unique papers. We then exclude tools that have already been analyzed, are not open-source (i.e., do not provide a specific code link), have not been updated in over six years, or are focused on specialized platforms (e.g., Emu5GNet \cite{RN1541}, EmuEdge \cite{RN1547}, or RaSim for Raspberry Pi \cite{RN1575}). However, all these filtered references are maintained in our GitHub repository to ensure comprehensive coverage.

\subsection{Comparison with Existing Surveys and Contribution}
Researchers in \cnc often resort to testing their methods using self-implemented scripts or tools due to a lack of standardized tools, which reveals a significant gap in the field. Earlier surveys, primarily focused on cloud computing, have not comprehensively addressed this issue. For instance, Wei et al. \cite{6486505} explore core engines (i.e., secondary development) and programming languages, categorizing them into software or hardware. Similarly, N. Mansouri et al. \cite{MANSOURI2020102144} argue that cloud simulation tools offer effective solutions for complex cloud computing environments, analyzing 33 tools. They highlight the absence of an ideal simulator that meets all user requirements, underscoring the need for further advancements.

In contrast to cloud computing, edge computing presents unique challenges, as demonstrated in \figurename~\ref{fig:relations}. These challenges include managing geo-distributed resources and computing across a continuum. It is widely acknowledged that cloud and fog (or edge) computing are sometimes interlinked, with applications or components deployable both on the cloud and the edge. While previous reviews such as those by Gupta et al. \cite{RN2687} focus predominantly on fog computing simulator and Wang et al. \cite{10.1007/978-3-031-30237-4_6} provide comparisons of MANET simulators, they lack in-depth consideration of edge-specific tools and scenarios. Furthermore, most existing reviews center on high-level aspects of networking or computing, without delving into the fusion of computation and networking specific to edge computing applications. 

In the domain of Peer-to-Peer networks, Shivangi et al. \cite{SURATI2017705} discover a preference for generic simulators over domain-specific or protocol-specific ones. They emphasize that high-level performance criteria, such as scalability, are crucial. In the Internet of Things (IoT) domain, quality characteristics like performance, reliability, and security are pivotal, as discussed in \cite{fi11030055}. It examines the qualities and metrics that simulators support in accordance with the ISO/IEC 25023 standard \cite{iso250232016} and reveals the challenge of constructing a framework that encompasses all necessary metrics. In the context of mobile ad-hoc networks (MANET), Jingzhi et al. \cite{10.1007/978-3-031-30237-4_6} studied simulators and emulators, comparing the running speed, scalability, and protocol support of five projects. They argue that large-scale MANET support should be further addressed.

Despite the recent emergence of numerous simulators, there is a notable absence of up-to-date surveys addressing both computing and networking. This survey seeks to bridge that gap, offering researchers a thorough guide to validate their methodologies in these rapidly evolving fields.

Our primary contributions are as follows:
\begin{itemize}
	\item We survey the most popular simulators and emulators in \cnc-related areas, classifying them into packet-level simulators, application-level simulators, and emulators. We compare them across five dimensions: Supported computing paradigms, simulated resources, supported performance metrics, resource management and utilization, and usability.
    \item For each dimension, we conduct a straightforward investigation to provide users (or readers) with a quick guide (a table) to select the appropriate tools for verifying their ideas.
    \item Based on our analysis, we propose five future directions: Integrating computing paradigms, processing data packets at the application level, simulating edge environments, supporting user-defined metrics, and enhancing scenario scripts and visualization.
\end{itemize}

This paper is organized into key sections, beginning with an introduction, research questions, and related works in the field of \cnc (Section \ref{sect_intro}). It then classifies simulators and emulators into three categories (Section \ref{sect_pack_app_level}), explores the core engines and supported computing paradigms (Section \ref{sect_computing_paradigms}), illustrates the simulated computing and communication resources (Section \ref{sect_resource_simulation}), lists the supported performance metrics (Section \ref{sect_performance_metrics}), investigates resource management and utilization (Section \ref{sect_resource_management}), and compares the usability between different simulators and emulators (Section \ref{sect_usability}). We conclude with future directions for research and development in edge computing simulation and emulation (Section \ref{sect_future_directions}), providing insights to guide researchers in selecting appropriate tools and advancing the field. Finally, we present our conclusions (Section \ref{sect_conclusion}).

\section{Overview and Paradigms of Simulators and Emulators}
\subsection{Simulator and Emulator Implementation Levels}
\label{sect_pack_app_level}
\begin{figure}
	\centering
	\includegraphics[width=\linewidth]{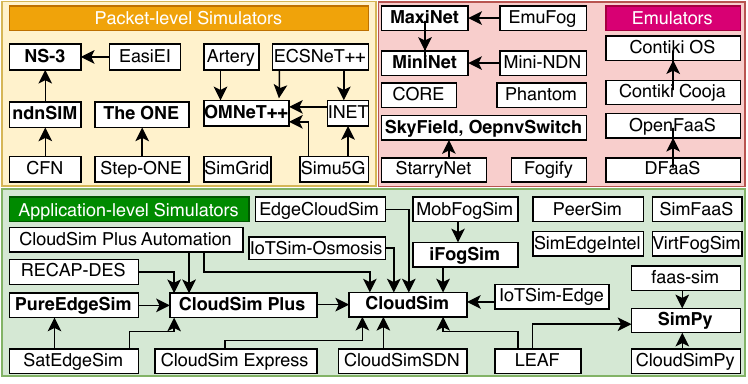}
	\caption{Categories and Inheritance of Simulators and Emulators}
	\label{fig:relations}
\end{figure}

Simulators and emulators operate at different levels of abstraction, from detailed packet-level simulations to high-level application modeling. The choice between them depends on the research objectives, as each type strikes a balance between accuracy, computational efficiency, and scalability. These tools can be broadly classified into three categories: Packet-level simulators, application-level simulators, and emulators, as illustrated in \figurename~\ref{fig:relations}. The engine behind each simulator or emulator plays a crucial role in determining its capabilities, scalability, and limitations. By including the core engines in our analysis, we can trace the tool’s functionality back to its framework, ensuring researchers select the most appropriate tool for their needs. Following the category order, \tablename~\ref{tab:compt-paradigms} further categorizes them into three distinct classes, separated by horizontal lines.

\textit{\textbf{Packet-level simulators}}, such as \NSthree \cite{nsnam}, \TheONE \cite{theone}, and \OMNeT \cite{omnetpp}, simulate the behavior of individual packets as they travel through the network. These simulators model detailed network events like packet arrival and departure at routers, making them useful for analyzing protocols and network performance at a fundamental level. A key advantage of packet-level simulators is their ability to measure precise performance metrics, such as latency and throughput, by tracking each packet \cite{Narayana2023MQL}. 

The engines powering these simulators, such as \NSthree and \OMNeT, are designed to support low-level network studies, focusing on detailed communication processes. These engines are often extended by tools like \CFN and \EasiEI to integrate in-network processing capabilities, allowing researchers to simulate complex computational paradigms while maintaining detailed control over communication. The core engines’ fine-grained control over packet behavior makes them invaluable for analyzing protocols in-depth and exploring novel networking strategies such as piggybacking \cite{9862944}.  However, because packet-level simulators process every packet individually, they become computationally expensive as network size increases, limiting their ability to handle large-scale networks.

In contrast, \textit{\textbf{application-level simulators}} take a higher-level approach, abstracting away the packet-level details to focus on broader system and application behaviors. This category includes both flow-level simulators like \faassim \cite{faas}, which treat groups of packets as single entities (flows), and simulators like \CloudSim \cite{cloudsim} and \iFogSim \cite{ifogsim2}, which focus on simulating the application-layer aspects of distributed computing systems. Flow-level simulators allow for more efficient simulations in large-scale networks or high-speed environments by reducing computational complexity. This makes them ideal for large networks where tracking each packet would be impractical \cite{Navaridas2019INRFlow}. Meanwhile, the latter category emphasizes the computational and resource management aspects at the application layer, modeling tasks such as job scheduling, resource allocation, and overall system performance. These simulators abstract the network layer, focusing instead on how applications run across distributed or cloud systems \cite{Li2020QoS-aware}. Tools like \CloudSim and \iFogSim could enable simulations with millions of nodes without the computational burden associated with packet-level simulators.

In addition, \CloudSim is the core engine behind most cloud and edge computing simulators, including \iFogSim and \EdgeCloudSim. It provides a flexible, extensible framework that allows researchers to model dynamic resource allocation and task scheduling in distributed infrastructures. The framework supports diverse computing environments, such as federated clouds, fog computing, and multi-tier edge architectures. By leveraging \CloudSim’s foundational capabilities, researchers can tailor the simulation to specific cloud or edge computing scenarios, while benefiting from a wide range of extensions that enhance its functionality.

When higher fidelity is required, \textit{\textbf{emulators}} come into play. Unlike simulators, emulators can run real application code or experiments on physical or virtualized infrastructures \cite{Zeng2019EmuEdge}. They support the entire networking and computing stack and can operate in controlled, realistic environments, making them suitable for debugging and system optimization. Emulators often leverage virtualization or multi-threading to simulate a distributed network on a single machine or a small cluster.

Emulators are powered by engines that support real-time interaction with actual hardware or network systems, enabling detailed experimentation with minimal abstraction. These engines are particularly useful for real-world testing and validation, as they can interface with live environments. Emulators like DFaaS \cite{dfaas} use trace data from real networks to provide high-fidelity feedback on performance, enabling researchers to optimize system configurations in a realistic setting.

\textit{Remark}: While we categorize the implementation levels into three types based on their core characteristics and primary functionalities, we admit that some simulators may incorporate overlapping features. For instance, some application-level simulators, like \CloudSim \cite{cloudsim} and \faassim \cite{faas}, are primarily designed for high-level task management and resource scheduling but may also utilize trace data from network simulators to simulate network latency or resource consumption. Similarly, packet-level simulators such as EasiEI, which focus on fine-grained network simulation, can use trace data to enrich their modeling of network behaviors. Emulators like DFaaS \cite{dfaas} might interface directly with real network environments and use trace data to debug and optimize system performance. 

To address these overlaps, our classification is primarily guided by the core purpose and most prominent use case of each tool. If a simulator’s main function is focused on capturing and analyzing network-level traffic, we classify it as a packet-level simulator. Conversely, if its primary focus is on high-level resource management and task scheduling, it is categorized as an application-level simulator, regardless of its use of network trace data. Emulators are classified based on their ability to interface with real networks or systems, providing more realistic experimental environments.

\subsection{Supported Computing Paradigms}
As summarized in \tablename~\ref{tab:compt-paradigms}, we categorize simulators and emulators based on the primary computing paradigms they support, such as cloud, fog, edge, and in-network computing, reflecting the diverse needs of modern distributed computing environments. 

\textbf{\textit{Cloud computing}} simulators model centralized, large-scale infrastructures, focusing on resource management, scheduling, and virtualization. Examples include \CloudSim \cite{cloudsim} and its derivatives like \CloudSimPlus \cite{cloudsimplus} and \CloudSimSDN \cite{cloudsimplus}, which allow the simulation of cloud environments, dynamic scaling, and network functions. These simulators are ideal for research into cloud resource allocation, federated clouds, and virtualized infrastructures.

\textbf{\textit{Edge \& fog computing}} simulators bring computation closer to the data source by simulating distributed environments across edge and fog nodes. \EdgeCloudSim \cite{edgecloudsim}, \iFogSim \cite{ifogsim2}, and \MobFogSim \cite{mobfogsim} are prominent tools in this category, focusing on task offloading, service migration, and real-time IoT applications. They support multi-tier architectures where both cloud and edge nodes interact, making them suitable for studies in latency reduction and dynamic task management.

\textbf{\textit{In-Network computing}} simulators extend traditional networking simulators to integrate computational tasks within the network itself, i.e., computing while transmitting. Tools like \CFN \cite{cfn} and \EasiEI \cite{easiei} build on \NSthree \cite{nsnam} and \OMNeT \cite{omnetpp} to simulate edge and in-network computing, supporting scenarios such as NFV (Network Function Virtualization) and service chaining. These simulators are particularly well-suited for research into the convergence of computing and communication paradigms, focusing on the optimization of resources at the network layer.

\textbf{\textit{Serverless \& mist computing}} simulators allow lightweight, distributed computation. Simulators such as \faassim \cite{faas} and \SimFaaS \cite{simfaas} support serverless function execution, modeling the dynamic behavior of functions without managing the underlying infrastructure. Mist computing, supported by tools like \PureEdgeSim \cite{pureedgesim}, focuses on ultra-lightweight tasks executed on resource-constrained devices at the very edge of the network.

\textbf{\textit{Specialized simulators \& emulators for unique scenarios}} address specialized computing environments. For instance, \SatEdgeSim  \cite{satedgesim} simulates satellite edge computing, while \SimufifthG \cite{simu5g} and \Artery \cite{artery} focus on vehicular communication and 5G networks. Emulators such as \MinNet \cite{mininet} and \MaxiNet \cite{maxinet} are designed to create realistic network environments, simulating large-scale networked systems and SDN. These tools are designed for scenarios requiring ultra-low latency, high mobility, or specific network architectures, making them ideal for dedicated applications such as V2X, satellite communications, and real-world SDN testing.

\begin{sidewaystable*}[htbp]
	\centering
	\renewcommand{\arraystretch}{1.2} 
	\caption{Computing Paradigms Supported in the Simulators/Emulators}
	\label{tab:compt-paradigms}
    \scriptsize 
	\begin{tabular}{p{4cm} | p{5cm} | p{4cm} | p{9.5cm}}
		\hline
		Simulator                                          & Paradigms                                                 & Core Engines                             & Highlights/Scenarios                                                            \\
        \hline 
		\NSthree \cite{nsnam}                              & Networking                                                & -                                 & Simulates network at packet-level                                               \\
		\OMNeT \cite{omnetpp} and INET                     & Networking                                                & -                                 & Simulates network at packet-level; excellent visualization                      \\
		\ndnSIM \cite{ndnsim}                              & Information-Centric Networking                            & \NSthree                          & Supports NDN protocols                                                          \\
		\TheONE \cite{theone}                              & Opportunistic Networking                                  & -                                 & Generates mobility traces, supports different routing protocols                 \\
        \Artery \cite{artery}                              & V2X                                                       & \OMNeT, SUMO, Vanetza            & Supports ETSI ITS-G5 protocols                                                  \\
		\SimufifthG \cite{simu5g,simulte}                  & Multi-access Edge, Cellular V2X                           & \OMNeT, INET                      & Supports 5G New Radio Networks                                                  \\
		\CFN \cite{cfn}                                    & Edge, In-network Computing                                & \ndnSIM                            & Supports NFaS                                                                   \\
		\EasiEI \cite{easiei}                              & Edge, Cloud Computing                                     & \NSthree                          & Supports various device representations, task scheduling policies               \\
		\ECSNeT \cite{ecsnet}                              & Edge, Distributed Stream Computing                        & \OMNeT, INET                      & Supports distributed stream processing applications                             \\
		\SimGrid \cite{simgrid}                            & Grid, P2P, Cloud, Fog Computing                           & -                                 & Offers rich APIs, practical abstraction models                                  \\
		\StepONE \cite{stepone}                            & IoT, Edge-Fog Computing                                   & \TheONE                           & Supports executing a set of processes (scenarios)                               \\
        \hline 
		\PeerSim \cite{peersim}                            & Networking (Peer-to-peer)                                 & -                                 & Extreme scalability and support for dynamicity                                  \\
        \CloudSim \cite{cloudsim}                          & (Federated) Cloud Computing                               & -                                 & Models (federated) cloud environment, scheduling, and allocating policies       \\
		\CloudSimPlus \cite{cloudsimplus}                  & (Federated) Cloud Computing                               & \CloudSim                         & Dynamically manages resources in runtime                                        \\
		\CloudSimExpress \cite{cloudsimexpress}            & Cloud Computing                                           & \CloudSim                         & Allows human-readable scenario scripts, dynamic injection of extensible modules \\
		\CloudSimPlusAutomation \cite{cloudsim-automation} & Cloud Computing                                           & \CloudSim, \CloudSimPlus          & Allows human-readable scenario scripts                                          \\
		\CloudSimSDN \cite{cloudsimsdn}                    & Cloud, Edge, In-network Computing                         & \CloudSim                         & Supports Edge Computing, NFV, and SFC support                                   \\
		\CloudSimPy \cite{cloudsimpy}                      & Cloud Computing                                           & SimPy                             & Supports deep learning frameworks, such as TensorFlow and PyTorch               \\
		\EdgeCloudSim \cite{edgecloudsim}                  & Fog, Cloud, Multi-access Edge Computing                   & \CloudSim                         & Supports WLAN/WAN, Mobility, and multi-tier edge computing architecture         \\
		\faassim \cite{faas}                               & Serverless Edge Computing                                 & SimPy                             & Replays traces from real-world platforms; supports function adaptations         \\
		\iFogSim \cite{ifogsim2}                           & Fog, (Mobile) Edge Computing                              & \CloudSim, \iFogSim                & Simulates the movement of IoT devices; supports microservices management        \\
		\IoTSimEdge \cite{iotsim-edge}                     & IoT, Edge Computing                                       & \CloudSim                         & Incorporates IoT and edge device behaviors, supports various network protocols  \\
		\IoTSimOsmosis \cite{osmosis}                      & IoT, Edge Computing                                       & \CloudSim                         & Supports edge-cloud heterogeneity and IoT application complexity                \\
		\LEAF \cite{leaf}                                  & Cloud, Edge Computing                                     & \CloudSimPlus, SimPy, NetworkX    & Emphasizes dynamic networks, power consumption modeling                         \\
		\MobFogSim \cite{mobfogsim}                        & Fog, (Mobile) Edge Computing                              & \iFogSim                          & Models device mobility and service migration                                    \\
		\PureEdgeSim \cite{pureedgesim}                    & Cloud, Edge, Mist Computing                               & \CloudSimPlus                     & Supports energy consumption analysis, dynamic workloads, AI algorithms          \\
		\RECAPDES \cite{recapdes}                          & Cloud to Edge Computing                                   & \CloudSimPlus                     & Simulates distributed data flow in a hierarchical architecture                  \\
		\SatEdgeSim \cite{satedgesim}                      & Satellite Edge Computing                                  & \CloudSimPlus, \PureEdgeSim       & Simulates satellite edge computing environment, supports high dynamic scenarios \\
		\SimFaaS \cite{simfaas}                            & Serverless Computing                                      & -                                 & Mimics public serverless computing platforms                                    \\
		\SimEdgeIntel \cite{simedgeintel}                  & Edge Computing, D2D                                       & -                                 & Simulates mobile networks for edge intelligence                                 \\
		\VirtFogSim \cite{virfogsim}                       & 5G Mobile-Edge-Cloud Computing                            & -                                 & Supports energy and delay performance optimization                              \\
		\YAFS \cite{yafs}                                  & Fog Computing                                             & -                                 & Supports network topology dynamics, placement, scheduling, and routing          \\
		\hline 
		\MaxiNet \cite{maxinet}                            & SDN                                                       & \MinNet                           & Emulates very large software-defined networks                                   \\
		\MinNet \cite{mininet}                             & SDN                                                       & -                                 & Creates a realistic virtual network on a single machine                         \\
		\Cooja \cite{cooja}                                & WSN                                                       & Contiki OS                        & Allows large WSN network emulation                                              \\
		\MiniNDN \cite{minindn}                            & Information-Centric Networking                            & \MinNet                           & Emulates testing and research on the NDN platform                               \\
		\Phantom \cite{shadow2}                            & Networking, Distributed Computing                         & \Shadow v1                        & Supports real networked applications; high scalability                          \\
		\StarryNet \cite{starrynet}                        & Satellite Networking                                      & SkyField, Docker, OpenvSwitch, tc & Mimics satellite dynamics and network behaviors                                 \\
		\CORE \cite{core}                                  & Edge, Cloud Computing                                     & -                                 & Builds a real computer network and runs it in real time                         \\
		\DFaaS \cite{dfaas}                                & (Federated) Edge Computing                                & OpenFaaS, libp2p                  & Balances traffic load across edge nodes; supports serverless function execution \\
		\EmuFog \cite{emufog}                              & Fog Computing                                             & \MaxiNet                          & Emulates fog computing infrastructures, including network topology generation   \\
		\Fogify \cite{fogify}                              & Fog, Edge Computing                                       & -                                 & Supports fog and edge computing typologies through Docker                       \\
        \hline
    \end{tabular}
\end{sidewaystable*}

\subsection{Lessons Learned}
As shown in \tablename~\ref{tab:compt-paradigms}, many simulators focus on multiple paradigms, reflecting the increasing convergence between computing and communication.
For instance, simulators like \CloudSimSDN, which extend traditional cloud simulations, support edge computing and in-network processing, providing a versatile platform for integrated computing and communication research. Meanwhile, tools like \SatEdgeSim are tailored for satellite edge computing, addressing the unique challenges of space-based communication.

Throughout our analysis of various simulators and emulators across different computing paradigms, we have gained several important insights.

\paragraph{Paradigm-Specific Tools Provide Deep but Narrow Support}
One of the key observations is that simulators designed for specific paradigms, such as packet-level network simulators or cloud computing simulators, tend to offer very detailed capabilities within their respective domains but often lack flexibility when applied to other paradigms. For example, \NSthree is highly detailed in simulating networking protocols but does not natively support cloud computing scenarios. This highlights the need for tools that offer more integration across paradigms to enable broader simulation capabilities.

\paragraph{Increasing Importance of Multi-Paradigm Simulators}
With the convergence of networking and computation in environments such as edge and fog computing, multi-paradigm simulators are becoming increasingly important. Tools like \CloudSimSDN and \EdgeCloudSim are valuable for research that spans both cloud and edge environments, offering a more integrated view of computing and networking. However, these tools are still evolving, and their ability to handle highly dynamic and distributed systems, especially in real-time or near-real-time applications, remains a challenge.

\paragraph{Emulation vs. Simulation Trade-offs}
Emulators such as \MinNet and \MaxiNet provide realistic environments by closely mimicking real-world network conditions. While this is useful for testing real-world applications and SDN, emulators can be resource-intensive and may lack the scalability of purely simulated environments. This trade-off between realism and scalability means that researchers must carefully select between simulation and emulation tools based on their specific needs, especially in large-scale experiments.

\section{Resource Simulation}
\label{sect_resource_simulation}
In this section, we explain the types of hardware resources supported by different simulators and emulators as classified in 	\tablename~\ref{tbl_resource_simu}. Specifically, we categorize the hardware representations into basic representations (simple computing components like CPUs and memory) and advanced representations (complex entities like cloud nodes and edge devices), focusing on both computing and communication resources. This categorization aims to provide a clearer understanding of the capabilities of each tool in representing the underlying infrastructure in CNC scenarios.

\subsection{Hardware Types}

For \textit{pure network simulation tools}, like \NSthree, \OMNeT, and \ndnSIM, they usually focus exclusively on network-level events. These packet-level simulators are primarily used to simulate network nodes and communication protocols, without involving detailed computational resources such as CPU or memory. They excel in modeling how data packets move across networks, which makes them suitable for studying communication protocols but less useful for simulating computational workloads.

For \textit{computing-capable simulation tools}, \textbf{\textit{basic representation}} means they focus on simulating simple computational components such as CPUs, memory, and storage. \faassim and \SimGrid belongs to this category, as it simulates basic components like GPUs, RAM, and CPUs.

\textbf{\textit{Advanced representation}} tools combine both computing and networking elements to provide a comprehensive model of system-wide interactions. \CloudSim and its variants, such as \CloudSimPlus, \iFogSim, and \EdgeCloudSim, use advanced node representations like VMs, computing nodes, and clusters for tasks like VM allocation, workload distribution, and mobility considerations in edge contexts. \SatEdgeSim supports satellite nodes to understand geo-distributed system impacts. \SimufifthG and \Artery extend \OMNeT to focus on vehicular and cellular systems, using constructs like roadside units , vehicular nodes, and 5G base stations for V2X scenarios. \MaxiNet and \MinNet represent physical and virtualized nodes, such as containers or VMs, for realistic SDN and NFV experiments.

Acturally, advanced representations can be further categorized into \textbf{\textit{general-purpose}} and \textbf{\textit{domain-specific}} types. Simulators like \CloudSim, \EasiEI, and \CloudSimPy use general-purpose advanced representations such as VMs, computing nodes, machines, and clusters. These representations are flexible and adaptable to various contexts. On the other hand, simulators derived from \CloudSim, such as \EdgeCloudSim, and tools like \SatEdgeSim, adopt domain-specific advanced representations that are more closely aligned with practical applications. They support nodes like edge datacenters, mobile devices, satellites, and gateways, which provide a more detailed and business-oriented perspective.

Therefore, different simulators and emulators support a wide range of hardware resources, from basic networking nodes in \NSthree and \MinNet to more advanced and complex environments like data centers and mobile devices in \CloudSim and \EdgeCloudSim. Basic representation simulators are best suited for modeling simple computational tasks, while advanced representation simulators provide a more holistic view of system-wide interactions involving both computing and networking elements. The primary distinction between simulators and emulators lies in their approach, with emulators providing a realistic environment and simulators relying on self-implemented network stacks.

\subsection{Communication Protocols}
Supported communication protocols, such as TCP/IP, are crucial when simulating network environments. Packet-level simulators, such as \NSthree and \Artery, primarily focus on the design and testing of communication protocols, offering detailed support for various networking standards. These simulators support the most protocols because protocols typically require detailed information stored in packets, such as headers and ports in TCP/UDP. 

Application-level simulators, such as \CloudSim and its variants, typically model network interactions as abstract delay matrices where each element represents the communication cost. Users of these simulators often need to build comprehensive mathematical models to accurately describe the effects introduced by the networking phase. This abstraction necessitates comprehensive mathematical models to accurately describe network effects.

Additionally, the classification of protocols varies among simulators. Some use general categories (e.g., WSN, WiFi, WLAN), while others, such as \NSthree and \Artery, use specific protocol standards. Collecting this information can be challenging, especially for application-level simulators, as it often requires a careful review of research papers and code repositories. Therefore, having a comprehensive list of supported protocols in the readme file of the simulator would significantly improve usability.

In summary, the support for communication protocols is critical, especially in packet-level simulators that require detailed packet information, such as TCP/UDP headers and ports. Application-level simulators often abstract these details, representing network interactions as delay matrices. This abstraction emphasizes the need for comprehensive documentation and clear readme files listing supported protocols to improve usability and accuracy in simulations.

\subsection{Resource Granularity and Representation}
When running an application or task, users need to specify the required resources. These specifications can include the number of nodes, CPUs, bandwidth, or even the number of floating-point operations (FLOPs). Consequently, the representation of resources can vary significantly.

As shown in \tablename~\ref{tbl_resource_simu}, simulators typically provide predefined resources in various forms, such as the number of nodes, CPU cores, processes, threads, or the number of instructions and storage size. For example, \CloudSim and its variants represent resources in terms of CPU cores, network delay matrix, power, and storage. On the other hand, \SimFaaS uses function execution time to denote resources. Additionally, certain simulators use basic parameters to represent resources more granularly. For example, \SimEdgeIntel uses transmission delay and noise power to model the network, providing a detailed view of network resource utilization.

Edge environments can indirectly affect resource performance. Factors such as temperature can impact computing phases, as noted in the environmental features (Env.) in \tablename~\ref{tbl_resource_simu}. Based on our investigation, we find that in edge computing, almost no simulator (\Artery is actually an ITS simulator). can support this feature. However, some conditions such as temperature and atmosphere can have an impact on computing or transmitting phases.

It is crucial to ensure that the granularity and representation of resources align with the specific needs of the application when selecting a simulator. The choice of simulator can significantly impact the accuracy and effectiveness of the verification process. A simulator that offers the right level of detail and appropriate resource representation will facilitate a more accurate and relevant simulation, leading to better insights and outcomes.

\begin{sidewaystable*}[htbp]
	\centering
	\caption{Resources Supported in the Simulators/Emulators}
	\renewcommand{\arraystretch}{1.2} 
	\footnotesize
	\label{tbl_resource_simu}
	\begin{tabular}{p{3cm} | p{5cm} | p{4cm} | p{9.5cm} | p{1cm}}
		\hline
		Simulator                                                                     & Hardware\footnotemark                       & Protocols                                               & Granularity \& Representation                                                                                    & Env.        \\
		\hline
		\NSthree \cite{nsnam}                                                         & simulated network                            & various (see the official \cite{ns3_model_library})    & network element (can be realistic)                                                                                             & NaF               \\
		\OMNeT                                                                        & simulated network                            & various (see the official \cite{omnetpp_models_tools}) & network element (can be realistic)                                                                                             & NaF               \\ 
		\ndnSIM \cite{ndnsim}                                                         & $\simeq$ \NSthree                            & NDN; $\simeq$ \NSthree                                 & $\simeq$ \NSthree                                                                                                              & NaF               \\
		\TheONE \cite{theone}                                                         & router                                       & DTN routing protocols                                  & radio (related to communication range and bit-rate), storage, energy (a budget)                                                & NaF               \\
		\Artery \cite{artery}                                                         & $\simeq$ \{OMNeT++, Vanetza\}                & ETSI ITS-G5 stack, IEEE 802.11                         & $\simeq$ OMNeT++                                                                                                               & weather, accident \\
		\SimufifthG \cite{simu5g, simulte}                                            & base stations, UEs; $\simeq$ \OMNeT          & 4G, 5G; $\simeq$ \{\OMNeT, INET\}                      & $\simeq$ realistic 4G and 5G parameters; $\simeq$ \{\OMNeT, INET\}                                                             & NaF               \\
		\CFN \cite{cfn}                                                               & computing resources;  $\simeq$ ndnSIM        & $\simeq$ ndnSIM                                        & function duration (an integer)                                                                                                 & NaF               \\
		\EasiEI \cite{easiei}                                                         & computing node; $\simeq$ \NSthree            & $\simeq$ \NSthree                                      & computing (\# resource units, resource pool); $\simeq$ \NSthree                                                                & NaF               \\
		\ECSNeT \cite{ecsnet}                                                         & edge device, cloud server                    & $\simeq$ \OMNeT and INET                               & \# CPU cores, CPU frequency, and \# threads/core, power (a number)                                                             & NaF               \\
		\SimGrid \cite{simgrid}                                                       & networks, CPUs, and disks                    & mathematical models for TCP/IP                         & network (bandwidth, latencies in numbers), computing (\# FLOPs), Disk (w/r rates)                                              & NaF               \\
		\StepONE \cite{stepone}                                                       & CPU cores                                    & $\simeq$ \TheONE                                       & \# CPU cores, CPU speed DMIPS + \TheONE                                                                                        & NaF               \\
        \hline
		\PeerSim \cite{peersim}                                                       & networking node                              & P2P (full list see the official \cite{peersim_extras}) & networking (delay matrix)                                                                                                      & NaF               \\
		\CloudSim, Plus  \cite{cloudsim, cloudsimplus}                                & physical and virtual machine, switch         & abstract model                                         & \# CPU cores, MIPS, network delay matrix, power and storage in numbers                                                               & NaF               \\
		CloudSim Express, 
        Plus Automation \cite{cloudsimexpress, cloudsim-automation}                   & $\simeq$ \CloudSim                           & $\simeq$ \CloudSim                                     & $\simeq$ \CloudSim                                                                                                             & NaF               \\
		\CloudSimSDN \cite{cloudsimsdn}                                               & edge datacenter, backbone, gateway           & NFV MANO architecture \cite{etsi2023opensource}        & $\simeq$ \CloudSim ; edge datacenter (physical machines); VNF (VMs)                                                            & NaF               \\
		\CloudSimPy \cite{cloudsimpy}                                                 & machine, cluster                             & trace-driven                                           & CPU, Memory, and Disk in numbers                                                                                               & NaF               \\
		\EdgeCloudSim \cite{edgecloudsim}                                             & mobile device, AP; $\simeq$ \CloudSim        & WLAN, WAN; $\simeq$ \CloudSim                          & $\simeq$ \CloudSim                                                                                                             & NaF               \\
		\faassim \cite{faas}                                                          & CPU, RAM, GPU, Network (traces)              & data flows (no specific protocols)                     & based on traces (a vector of resource utilization metrics)                                                                     & NaF               \\
		\iFogSim \cite{ifogsim2}                                                      & fog device, actuator, sensor                 & $\simeq$ \CloudSim                                     & $\simeq$ \CloudSim                                                                                                             & NaF               \\
		\IoTSimEdge \cite{iotsim-edge}                                                & sensor, edge datacenter                      & IoT protocols; $\simeq$ \CloudSim                      & $\simeq$ \CloudSim                                                                                                             & NaF               \\
		\IoTSimOsmosis \cite{osmosis}                                                 & IoT, edge datacenter, cloud datacenter       & SDN; $\simeq$ \IoTSimEdge                              & $\simeq$ \CloudSim                                                                                                             & NaF               \\
		\LEAF \cite{leaf}                                                             & datacenter and sensor                        & abstract model (based on graphs)                       & a weighted, directed multigraph                                                                                                & NaF               \\
		\MobFogSim \cite{mobfogsim}                                                   & $\simeq$ \iFogSim                            & $\simeq$ \iFogSim                                      & $\simeq$ \iFogSim                                                                                                              & NaF               \\
		\PureEdgeSim \cite{pureedgesim}                                               & datacenter, server, end device               & ethernet, cellular, WiFi                               & networking delay matrix,  \# CPU Cores, MIPS/core, storage (in Mb), energy (related to task's MIPS and transmitting data size) & NaF               \\
		\RECAPDES \cite{recapdes}                                                     & switch, router, computing node               & $\simeq$ \CloudSimPlus                                 & $\simeq$ \CloudSimPlus                                                                                                         & NaF               \\
		\SatEdgeSim \cite{satedgesim}                                                 & cloud, edge, mist node, such as satellite    & $\simeq$ \PureEdgeSim                                  & $\simeq$ \PureEdgeSim                                                                                                          & NaF               \\
		\SimFaaS \cite{simfaas}                                                       & service-related time instead                 & NaF                                                    & function execution time (in numbers)                                                                                           & NaF               \\
		\SimEdgeIntel \cite{simedgeintel}                                             & cloud server, base station and mobile device & D2D link                                               & storage (\# caching size), network (transmission delay, noise power, etc.)                                                     & NaF               \\
		\VirtFogSim \cite{virfogsim}                                                  & cloud, fog, mobile devices                   & TCP/IP                                                 & the processing frequency, and the throughput                                                                                   & NaF               \\
		\YAFS \cite{yafs}                                                             & fog nodes                                    & abstract model (at application-level)                  & instructions per time, memory capacity, bandwidth (in bytes), link propagation delay, and other user-defined resources         & NaF               \\
		\hline
		\MaxiNet \cite{maxinet}                                                       & realistic  network                           & $\simeq$ realistic network stack and SDN               & container node                                                                                                                 & NaF               \\
		\MinNet \cite{mininet}                                                        & realistic network                            & $\simeq$ realistic network stack and SDN               & Process                                                                                                                        & NaF               \\
        \Cooja \cite{cooja}                                                           & sensor                                       & WSN, 6TiSCH, 6LoWPAN                                   & CPU, memory in physical form                                                                                                   & NaF               \\ 
		\MiniNDN \cite{minindn}                                                       & $\simeq$ \MinNet                             & NDN; $\simeq$ \MinNet                                  & $\simeq$ \MinNet                                                                                                               & NaF               \\
		\Phantom \cite{shadow2}                                                       & networking node                              & $\simeq$ realistic network stack                       & Process                                                                                                                        & NaF               \\
		\StarryNet \cite{starrynet}                                                   & satellite, ground-station, terminal          & realistic network stack                                & container (users can configure the CPU capability)                                                                             & NaF               \\
		\CORE \cite{core}                                                             & realistic network                            & $\simeq$ realistic network stack                       & container                                                                                                                      & NaF               \\
		\DFaaS \cite{dfaas}                                                           & realistic network or traces                  & TCP/HTTP                                               & container, link                                                                                                                & NaF               \\
		\EmuFog \cite{emufog}                                                         & realistic network                            & $\simeq$ \MaxiNet                                      & $\simeq$ \MaxiNet                                                                                                              & NaF               \\
		\Fogify \cite{fogify}                                                         & realistic network                            & realistic network stack                                & CPU (cores and cycles), network (as VXLAN), memory and disk                                                                    & NaF               \\
		\hline
	\end{tabular}
	\begin{minipage}{24cm}
		\smallskip
		\textsuperscript{1}``$\simeq$'' means the features are the same with or extended from.  \\
        \textsuperscript{2} NaF: Not a feature or not mentioned. \\
  
	\end{minipage}
\end{sidewaystable*}

\section{Performance Metrics}
\label{sect_performance_metrics}
Understanding performance metrics across various simulators and emulators is crucial for researchers to make decisions about the tools they use for optimizing and evaluating network and service performance. We classify performance metrics into task-wide and system-wide. Task-wide metrics relate to sources used for optimizing user-specific tasks, while system-wide metrics relate to the simulator/emulator performance during experiments. In this section, we survey task (including basic and custom) and system-wide metrics that might be useful for accelerating the evaluation phases.

\subsection{Supported Basic Metrics}
Most simulators measure common network and service performance metrics such as latency, energy consumption, and resource usage. As shown in \tablename~\ref{tbl_metrics}, packet-level simulators typically support more fundamental and fine-granularity measurements (e.g., jitter, error rate, and hop-by-hop latency), whereas application-level simulators support higher-level metrics (e.g., service execution time, service migration delay, and end-to-end latency). Additionally, statistical metrics such as average instance lifespan and service execution success probabilities can also be used, as seen in tools like \SimFaaS.

In edge computing, particularly in mobile environments, task offloading is a critical scenario. However, while mobile features are common in packet-level simulators, related measurements, such as migration delay, are still rare.

For emulators, basic metrics can be directly obtained using well-developed tools such as \texttt{top} and \texttt{iPerf} on Linux. However, this approach may involve significant engineering effort and high costs.

\subsection{Custom Metrics or Interfaces}
The ability to define custom metrics is crucial for users with specific needs. For example, service reliability, representing the probability that total latencies are below a given threshold, can be composed of transmitting latency and computing time \cite{remr}. Implementing such custom metrics often requires interfaces or tools for analyzing databases and log files.

According to our investigation, as shown in \tablename~\ref{tbl_metrics}, not all simulators or emulators provide this feature. About eight simulators have built-in classes/interfaces to support custom metrics, enhancing their flexibility and applicability for diverse use cases. However, the lack of custom metrics support can limit the usefulness of a simulator or emulator for specific research requirements.

For emulators, only two support custom metrics: \StarryNet, which supports network connectivity measurement, and \Fogify, which allows users to define their measurements at the application level.

In summary, while some simulators and emulators offer built-in support for custom metrics through interfaces and tools, this feature is not universally available. The capability to define or calculate custom metrics is essential for tailoring simulations to specific research needs, and its absence can constrain the adaptability and effectiveness of simulation tools.

\subsection{System-wide Performance}
Simulators or emulators that support numerous networked nodes and scheduling policies consume significant host resources. Therefore, evaluating this metric is critical when choosing the most suitable option among various candidates.

As shown in \tablename~\ref{tbl_metrics}, application-level simulators can support a large number of nodes, such as up to $10^7$ nodes for \PeerSim and $1$ million nodes for \CloudSim and its variants. This capability is due to the abstraction of data packet processing logic, which differs from the more detailed packet-level simulators. Emulators typically support smaller scales but can be extended by running them over multiple machines. Examples include \Phantom and \StarryNet, which can scale their capabilities through distributed execution.

Moreover, although system-wide performance is correlated with users' tasks, not all simulators list this metric, particularly concerning scalability. The maximum number of supported nodes is not always clearly stated, making it challenging to assess the scalability of some simulators. 

In summary, when selecting simulators or emulators, it is crucial to consider the host resources they consume, their ability to support a large number of nodes, and their scalability. Application-level simulators generally support larger scales due to their abstraction of packet processing, while emulators can achieve scalability through distributed execution. However, the absence of explicit scalability metrics in some simulators necessitates careful evaluation based on specific user requirements.

\begin{sidewaystable*}
	\centering
	\scriptsize 
	\renewcommand{\arraystretch}{1.2} 
	\caption{Metrics}
	\label{tbl_metrics}
	\begin{tabular}{p{4cm} p{10cm} p{2.2cm} p{7cm}}
		\hline
		Simulator                                                                                   & Basic                                                                                                                                   & Custom (Interfaces)              & System-wide                                                   \\
		\hline
		\NSthree \cite{nsnam}                                                                       & bandwidth, latency, throughput, jitter, packet loss rate, energy consumption, etc.                                                      & Tracing System                   & 20k+ \cite{9921600}                                              \\
		\OMNeT \cite{omnetpp}                                                                       & bandwidth, latency, throughput, jitter, packet loss rate, energy consumption, etc.                                                      & \tt{cStatistic}                  & $\sim$5k \cite{9921600}                                          \\
		\ndnSIM \cite{ndnsim}                                                                       & bandwidth, latency,  packet loss; $\simeq$ \NSthree                                                                                     & NaF                              & NaF                                                              \\
		\TheONE \cite{theone}                                                                       & energy level, inter-contact times, messages, message delivery ratio, message life                                                       & NaF                              & NaF                                                              \\
		\Artery \cite{artery}                                                                       & NaF                                                                                                                                     & NaF                              & NaF                                                              \\
		\SimufifthG  \cite{simu5g,simulte}                                                          & latency, error rate, and metrics from \OMNeT, INET                                                                                     & NaF                              & NaF                                                              \\
		\CFN \cite{cfn}                                                                             & inherited from ndnSIM                                                                                                                   & NaF                              & scalability                                                      \\
		\EasiEI \cite{easiei}                                                                       & resource usage, latency (end-to-end, hop-by-hop); $\simeq$ NS-3                                                                         & $\simeq$ \NSthree                & NaF                                                              \\
		\ECSNeT \cite{ecsnet}                                                                       & delay (processing, transmitting, total), average power consumption; $\simeq$ \OMNeT                                                     & $\simeq$ \NSthree                & NaF                                                              \\
		\SimGrid \cite{simgrid}                                                                     & task execution time, resource usage, communication overhead, system reliability and robustness,  energy consumption                     & NaF                              & $\sim$2m processors, 16GB RAM                                    \\
		\StepONE \cite{theone}                                                                      & costs and budgets per CPU; $\simeq$\TheONE                                                                                              & \texttt{CostHelper}              & NaF                                                              \\
		\hline
		\PeerSim \cite{peersim}                                                                     & latency                                                                                                                                 & NaF                              & $>$ $10^7$ (cycle-based) and $10^5$ (event-based) nodes, 4Gb RAM \\
		\CloudSim \cite{cloudsim}, \CloudSimPlus \cite{cloudsimplus}                                & latency (execution, transmission), resource usage, power consumption                                                                    & NaF                              & 1m nodes, 2.27GHz Intel Core Duo x 2, 16GB RAM                     \\
		\CloudSimExpress \cite{cloudsimexpress}, \CloudSimPlusAutomation \cite{cloudsim-automation} & $\simeq$ \CloudSim                                                                                                                      & NaF                              & $\simeq$ \CloudSim; code complexity reduction                    \\
		\CloudSimSDN \cite{cloudsimsdn}                                                             & VNF throughput;  link capacity;  $\simeq$ \CloudSim                                                                                     & NaF                              & $\simeq$ \CloudSim                                               \\
		\CloudSimPy \cite{cloudsimpy}                                                               & cluster and machine state                                                                                                               & NaF                              & NaF                                                              \\
		\EdgeCloudSim \cite{edgecloudsim}                                                           & Task Failure Rate (includes mobility-related failures), WLAN/WAN delay; $\simeq$ \CloudSim                                              & NaF                              & $\simeq$ \CloudSim                                               \\
		\faassim \cite{faas}                                                                        & function execution time, resource usage,  latency                                                                                       & NaF                              & CPU and RAM usage                                                \\
		\iFogSim \cite{ifogsim2}                                                                    & mobility-related(migration delay, energy consumption, etc.); $\simeq$ \CloudSim                                                         & NaF                              & resources usage, 2.33GHz Intel Core 2 Duo, 2 GBRAM               \\
		\IoTSimEdge \cite{iotsim-edge}                                                              & mobility related metrics, migration delay, energy consumption; $\simeq$ \CloudSim                                                       & NaF                              & 19.14 MB RAM for 50 IoT devices (no chip specified)              \\
		\IoTSimOsmosis \cite{osmosis}                                                               & $\simeq$ \CloudSim                                                                                                                      & NaF                              & 520 devices, 200MB RAM (no chip specified)                       \\
		\LEAF \cite{leaf}                                                                           & detailed energy consumption                                                                                                             & NaF                              & 46.5k taxis ($sim$ 330k tasks), 1.4GHz Intel Single Core         \\
		\MobFogSim \cite{mobfogsim}                                                                 & $\simeq$ \iFogSim; mobility related metrics such as moving speed                                                                       & NaF                              & NaF                                                              \\
		\PureEdgeSim \cite{pureedgesim}                                                             & delays, energy consumption, resources utilization, and tasks success rate                                                               & NaF                              & 10k+ devices, a single Intel Core i7-8550U                       \\
		\RECAPDES \cite{recapdes}                                                                   & CPU, memory, storage, and power utilization, bandwidth, end-to-end response time                                                        & \texttt{Utilisation}             & NaF                                                              \\
		\SatEdgeSim \cite{satedgesim}                                                               & delays, energy consumption, and task success rate                                                                                       & NaF                              & NaF                                                              \\
		\SimFaaS \cite{simfaas}                                                                     & probabilities related to service execution, average instance life span, average server count, average running count, average idle count & \texttt{SimProcess}              & NaF                                                              \\
		\SimEdgeIntel \cite{simedgeintel}                                                           & cache hit rate, content popularity, delivery latency                                                                                    & NaF                              & NaF                                                              \\
		\VirtFogSim \cite{virfogsim}                                                                & computing frequency, transport bit rate, energy consumption                                                                             & NaF                              & NaF                                                              \\
		\YAFS \cite{yafs}                                                                           & average response time, link latency, resource utilization, etc.                                                                         & \texttt{Stats}, \texttt{Metrics} & NaF                                                              \\
		\hline
		\MaxiNet \cite{maxinet}                                                                     & CPU utilization, memory consumption, and network usage                                                                                  & NaF                              & the speedups for actual experiments are more than linear         \\
		\MinNet \cite{mininet}                                                                      & bandwidth, latency, packet loss, queue type (FIFO, SFQ, TBF), and CPU limits                                                            & NaF                              & $\sim$1k nodes, 2.4 GHz Intel Core 2 Duo, 6GB RAM                \\
		\Cooja \cite{cooja}                                                                         & packet transmission rate, end-to-end latency, network throughput, power consumption, signal strength, communication range, etc.         & NaF                              & NaF                                                              \\
		\MiniNDN \cite{minindn}                                                                     & $\simeq$ \MinNet                                                                                                                        & NaF                              & NaF                                                              \\
		\Phantom \cite{shadow2}                                                                     & latencies, bandwidth, CPU and RAM usage, and other statistics at packet level                                                           & NaF                              & 64k nodes, 2.6 GHz Intel Xeon 2x14 cores,  256GB RAM             \\
		\StarryNet \cite{starrynet}                                                                 & CPU/memory/bandwidth usage                                                                                                              & connectivity                     & 4408 nodes, 7 machines (two Intel Xeon, 8*32GB RAM)              \\
		\CORE \cite{core}                                                                           & bandwidth, delay, and loss                                                                                                              & NaF                              & $\sim$100 nodes, 3.0GHz single-CPU Xeon, 2GB RAM                 \\
		\DFaaS \cite{dfaas}                                                                         & latency, workload states                                                                                                                & NaF                              & NaF                                                              \\
		\EmuFog \cite{emufog}                                                                       & $\simeq$ \MaxiNet                                                                                                                       & NaF                              & NaF                                                              \\
		\Fogify \cite{fogify}                                                                       & CPU time, memory usage, disk I/O, network traffic                                                                                       & application-level                & CPU and Network usage, real service deployment                   \\
		\hline
	\end{tabular}
	\begin{minipage}{24cm}
		\smallskip
	\end{minipage}
\end{sidewaystable*}

\section{Resource Management and Utilization}
\label{sect_resource_management}
We further investigate resource managements and utilization details in \tablename~\ref{tbl_resource_mgmt}. Different from hardware or roles in \tablename~\ref{tbl_resource_simu}, we further list the resources in a fine-grained, such as CPU, memory, because a node in \cnc usually contains both computing and communicating capability. 
\tablename~\ref{tbl_resource_mgmt} includes three categories: Resources allocated (i.e., Allocation what), scheduling policies, and cost management or optimization target the simulator supported.

\subsection{Resource Allocation Types}
Resource allocation types can be classified into node-level and unit-level, as detailed in \tablename~\ref{tbl_resource_mgmt}. Node-level allocation involves self-consistent nodes, typically using virtualization techniques like Docker, VM, and Process. Most emulators follow this approach, allocating isolated resources to each node and running real applications within these virtualized environments.

For unit-level allocation, resources are further broken down into bandwidth (BW), computation (CPU), memory, and energy. Simulators like \SimFaaS use instructions and service functions, allowing experiments or trace replays at the function level. This implies each function can be treated as a node, catering to serverless computing needs.

\subsection{Scheduling Policies}
Scheduling policies are crucial in both networking and computing simulators, as shown in \tablename~\ref{tbl_resource_mgmt}. Most simulators support conventional scheduling policies, such as First-Come First-Serve (FCFS) and Round Robin. Meanwhile, some simulators also support
advanced scheduling methods, such as reinforcement learning-based (\CloudSimPy) and genetic algorithm-based (\VirtFogSim).

Scheduling in \cnc involves a trade-off between computation and communication resources. Application-level simulators typically support a wide range of policies integrated with networking at the application level, allowing diverse user requirements and optimization indicators. Mobility is another critical factor, often involving energy and location considerations. Although mobility is well-supported in packet-level simulators, integrating it with computing is less common.

Meanwhile, built-in scheduling policies may not always meet user requirements, making user-defined features important. This customization capability is highlighted in \tablename~\ref{tbl_resource_mgmt}.

We also find that some scheduling policies are overlapped, particularly in packet-level simulators. Queuing theory can be used for timely packet delivery, with policies like FCFS and priority-based scheduling. Similarly, scheduling tasks or services can mirror packet scheduling, suggesting a need for a common scheduling library in packet-level simulators.

\subsection{Cost Management}
We refer cost as the time, energy, or any other defined indicators consumed by running the scheduled service/task deployment. Actually, due to diverse user requirements,, having a uniformed optimizing target is unrealistic. However, to give readers a helpful guide, we also list the built-in cost management in \tablename~\ref{tbl_resource_mgmt}. We can see that as the area becomes more specific, the ways of cost management emphasized are also constrained. For example, \CloudSim supports max resource utilization, while \iFogSim based on that optimizes the cost of task migration in mobile scenario, \CloudSimSDN handles the shortest path in SDN, and \MobFogSim focuses on proximity to reduce latencies.  However, fundamental cores such as \NSthree, \OMNeT, and emulators are actually without optimizing target. Therefore, find an simulator that can quickly support user-defined cost management APIs can also be helpful.

\begin{sidewaystable*}
	\centering
	\footnotesize
	\renewcommand{\arraystretch}{1.2} 
\caption{Resource Management and Utilization}
\label{tbl_resource_mgmt}
    \begin{tabular}{p{3cm} p{4cm} p{9cm} p{7cm}}
        \hline
        Simulator &  Allocation What &  Scheduling Policies & Cost Management (Or Optimization Target)\footnotemark \\
        \hline
        \NSthree \cite{nsnam} & network nodes, BW, energy & simulator events: calendar queue, heap sort, double linked-list, etc.; user-defined & user-defined \\
        \OMNeT \cite{omnetpp} & network nodes, BW, energy & real-time or sequential scheduling (simulator events), user-defined & user-defined \\
        \ndnSIM \cite{ndnsim} & $\simeq$ \NSthree & $\simeq$ \NSthree & $\simeq$ \NSthree \\
        \TheONE \cite{theone} & network nodes & FIFO, random, etc. (for messages) & NaF \\
        \Artery \cite{artery}& vehicles (OMNeT++ modules) & Data replaying & NaF \\
        \SimufifthG \cite{simu5g,simulte} & CPU, memory, disk; $\simeq$ \OMNeT & max C/I, proportional fair, round robin, etc. & user-defined \\
        \CFN \cite{cfn} & computation ; $\simeq$ ndnSIM & based on name location and resource availability & first fit \\
        \EasiEI \cite{easiei} & resource units &  FCFS, round robin, high-priority, user-defined & user-defined \\
        \ECSNeT \cite{ecsnet} & CPU Core and thread & round robin, predefined plan, user-defined & min latencies, max throughput, optimize energy cost \\
        \SimGrid \cite{simgrid} & CPU, network, storage & performance-based, load balancing, data-aware, energy-efficient, priority & user-defined \\
        \StepONE \cite{stepone} & CPU instructions, BW &  user-defined & custom, such as place processes among fog and cloud \\
       \hline
        \PeerSim \cite{peersim} & network elements & heap sort based on scheduled time (simulator events) & NaF \\
        \CloudSim, Plus \cite{cloudsim, cloudsimplus} & VMs, memory, storage, BW & FCFS, space-shared, time-shared, user-defined &  max resource utilization, correlation; random, user-defined \\
        \CloudSimExpress, Plus Automation \cite{cloudsimexpress, cloudsim-automation}& $\simeq$ \CloudSim & $\simeq$ \CloudSim & $\simeq$ \CloudSim \\
        \CloudSimSDN \cite{cloudsimsdn} & $\simeq$ \CloudSim &  VM/VNF/Container: first fit, most full; network: first met, random, modulo operation-based, most remaining capacity; $\simeq$ \CloudSim & shortest path; $\simeq$ \CloudSim \\
        \CloudSimPy \cite{cloudsimpy} & resource usage ratios & first fit, reinforcement learning, random & min make-span; average slowdown, completion of jobs \\
        \EdgeCloudSim \cite{edgecloudsim} & $\simeq$ \CloudSim & least loaded; $\simeq$ \CloudSim & min latency, energy consumption; balance load; $\simeq$ \CloudSim \\
        \faassim \cite{faas} & memory, computation, network & Skippy-based (a scheduling system \cite{RAUSCH2021259} ); fair and max-min (max bottleneck) flow allocation; user-defined & min function execution time, max resource usage \\
        \iFogSim \cite{ifogsim2} & $\simeq$ \CloudSim & mobility-aware; $\simeq$ \CloudSim & min migration delay, user-defined \\
        \IoTSimEdge \cite{iotsim-edge}& $\simeq$ \CloudSim & $\simeq$ \CloudSim & user-defined \\
        \IoTSimOsmosis \cite{osmosis} & computation, network, battery & network: shortest path, shortest path \& max BW; VM: least used resources, time-shared; $\simeq$ \CloudSim & user-defined \\
        \Cooja \cite{cooja} & memory, computation & priority-based, event-based ($\simeq$ Contiki OS) & various ($\simeq$ Contiki OS)\\
        \LEAF \cite{leaf} & memory, computation, network, energy & energy-aware & reduce static power usage by min \# of running nodes \\
        \MobFogSim \cite{mobfogsim} & $\simeq$ \CloudSim & migration based on network and location-aware metrics, user-defined & reducing latency, ensuring proximity, optimizing QoS \\
        \PureEdgeSim \cite{pureedgesim} & $\simeq$ \CloudSimPlus & round robin, trade-off (computation \& communication), user-defined & user-defined \\
        \RECAPDES \cite{recapdes} & $\simeq$ \CloudSimPlus & used to evaluate different deployments & in terms of cost, energy, resource allocation, and utilization \\
        \SatEdgeSim \cite{satedgesim} & $\simeq$ \PureEdgeSim & round robin, trade-off (transmission distance, energy consumption, task parallelism, CPU processing time), traditional polling, user-defined & user-defined \\
        \SimFaaS \cite{simfaas} & service instance & scale-per-request, concurrency value scaling; metrics-based scaling & min response time, reduce cold-starts, improve resource efficiency, user-defined \\
        \SimEdgeIntel \cite{simedgeintel} & cache &  FIFO, greedy, Knapsack, least-used, least recently used, etc. & max cache hit rates and min average transmission delay \\
        \VirtFogSim \cite{virfogsim} & processing frequency, throughput &  genetic task allocation, only mobile/fog/cloud allocation, exhaustive search & min total consumed energy \\
        \YAFS \cite{yafs} & CPU, BW & cluster placement (select the cheapest cluster), edge placement & random, the shortest path,  user defined \\
        \hline
        \MaxiNet \cite{maxinet} & virtual node & user-defined & user-defined \\
        \MinNet \cite{mininet} & host, switch & NaF & NaF \\
        \MiniNDN \cite{minindn} & $\simeq$ \MinNet & $\simeq$ \MinNet & $\simeq$ \MinNet \\
        \Phantom \cite{shadow2} & virtual node & CPU affinity, work stealing & NaF \\
        \StarryNet \cite{starrynet} & virtual node (docker) & shortest distance, longest remaining visible time & NaF \\
        \CORE \cite{core} & virtual node & NaF & NaF \\
        \DFaaS \cite{dfaas} & virtual node (emulation); CPU, RAM ratios (simulation) & equal, margin, and power saving strategies (emulation); empirical random, weight-based strategies (simulation); user-defined strategies & load balancing \\
        \EmuFog \cite{emufog} & $\simeq$ \MaxiNet & latency-based, user-defined & keep a latency bound \\
        \Fogify \cite{fogify} & virtual node (docker) & NaF & user-defined \\
        \hline
    \end{tabular}
\end{sidewaystable*}

\section{Supported Features and Popularity}
\label{sect_usability}
In simulators and emulators, various layers and components operate together to ensure the comprehensiveness and functionality of the simulation environment. We compare their types, programming languages, and components in \tablename~\ref{tbl_usability}. In addition, we also provide project statistics related to their academic impact and codebase metrics.

We first classify them into discrete-event simulators (DES)\cite{robinson2014simulation}, emulators (Emu), and hybrid (supporting both simulation and emulation). DES is the most popular type, capable of supporting middle to large-scale network topologies. Packet-level simulators are usually implemented in C/C++, while application-level simulators are often written in Java. This distinction may be due to the need for more basic processing and resource efficiency in handling data packets.

Then, we organize the modular design of these tools into eight layers, as shown in \tablename~\ref{tbl_usability}:

\begin{itemize}
	\item Service or Application Layer (\textbf{SEV}): This layer runs services or applications within the simulator, such as HTTP servers, databases, or any user-defined services.
	\item Facility Layer (\textbf{FAC}): Acts as an intermediary between the service layer and the communication layer, managing the distribution of service requests and the transmission and routing of data.
	\item Hardware Layer (\textbf{HW}): Enables the simulation of diverse hardware configurations, including network devices like switches and routers, as well as various computing and communication resources, thus emulating hardware heterogeneity in real network environments.
	\item Network Layer (\textbf{NET}): Responsible for the creation and management of network topologies, defining network links that interconnect nodes, and configuring network attributes such as bandwidth and latency.
	\item Configuration File Module (\textbf{CONFIG}): Allows users to define and load simulation configurations, including network topology, node, and service configurations, stored as text files for convenient editing and reuse.
	\item Scenario Script Module (\textbf{SCEN}): Permits users to define specific scenarios for simulations, including initiating network services, generating traffic, and executing network commands, thereby automating the simulation process. This is particularly significant for complex simulation scenarios such as smart cities.
	\item Logging Component (\textbf{LOG}): Records various events and data during the simulation, including system logs and network performance metrics, which are invaluable for analyzing and debugging network behavior.
	\item Visualization Module (\textbf{VIS}): Provides a graphical interface to display network topologies and simulation results, aiding users in intuitively understanding the network structure and performance.
\end{itemize}

As depicted in \tablename~\ref{tbl_usability}, packet-level simulators generally have a well-developed modular design. In contrast, application-level simulators may need further development in scenario scripting and visualization support. The challenge with visualization in application-level simulators lies in their inability to simulate data packets during transmission, which can limit their network design and visualization capabilities.

In summary, the usability of simulators and emulators is enhanced by their modular designs, which facilitate comprehensive and functional simulation environments. The choice of programming language and the modular design, including support for scenario scripts and visualization, play crucial roles in determining the suitability of a simulator or emulator for specific applications.

We believe these adjustments, along with the detailed metrics in Section V, will provide a clearer and more accurate representation of the simulators’ capabilities. We appreciate your valuable comments and look forward to any further suggestions you may have.

\begin{table*}
	\centering
	\renewcommand{\arraystretch}{1.2} 
	\caption{Supported Features and Popularity}
	\label{tbl_usability}
	\begin{tabular}{p{3.8cm} p{0.6cm} p{1cm} >{\centering\arraybackslash}p{0.25cm} >{\centering\arraybackslash}p{0.25cm}  >{\centering\arraybackslash}p{0.25cm} >{\centering\arraybackslash}p{0.25cm} >{\centering\arraybackslash}p{0.25cm} >{\centering\arraybackslash}p{0.25cm} >{\centering\arraybackslash}p{0.25cm} >{\centering\arraybackslash}p{0.25cm} p{5.8cm}}
		\hline
		Simulator                                          & Types\footnotemark                                                     & PL\footnotemark    & SEV    & FAC    & HW     & NET    & COF    & SCN    & LOG    & VIS\footnotemark    & Statistics\footnotemark                                                                         \\
		\hline
		\NSthree \cite{nsnam}                              & DES                                                                    & \CPP \Python       & \cmark & \cmark & \cmark & \cmark & \cmark & \cmark & \cmark & \cmark & 1.4k/GPL/S(208),F(528), W(35), C(23)/2008                                                       \\
		\TheONE \cite{theone}                              & DES                                                                    & \Java              & \cmark & \cmark & \xmark & \cmark & \cmark & \xmark & \cmark & \cmark & 2983/GPL/S(200), F(195), W(23), C(10)/2009                                                      \\
		\Artery \cite{artery}                              & DES                                                                    & \CPP               & \cmark & \cmark & \xmark & \xmark & \cmark & \cmark & \cmark & \cmark & 33/GPL/S(182), F(122), W(27), C(18)/2019                                                        \\
		\OMNeT \cite{omnetpp}                              & DES                                                                    & \CPP \Java         & \cmark & \cmark & \cmark & \cmark & \cmark & \cmark & \cmark & \cmark & 2.8k/S(546),F(141), W(30), C(12)/1997                                                           \\
		\ndnSIM \cite{ndnsim}                              & DES                                                                    & \CPP \Python       & \cmark & \cmark & \xmark & \cmark & \cmark & \cmark & \cmark & \cmark & 360/GPL/S(107),F(163), W(35), C(36)/2017                                                        \\
		\SimufifthG \cite{simu5g,simulte}                  & Hybrid                                                                 & \CPP               & \cmark & \cmark & \cmark & \cmark & \cmark & \cmark & \cmark & \cmark & 244/LGPL/S(137), F(71), W(15), C(6)/2020                                                        \\
		\CFN \cite{cfn}                                    & DES                                                                    & \CPP               & \cmark & \cmark & \xmark & \cmark & \cmark & \cmark & \cmark & \cmark & 96/GPL/S(6), F(4), W(3), C(1)/2019                                                              \\
		\EasiEI \cite{easiei}                              & DES                                                                    & \CPP \Python       & \cmark & \cmark & \cmark & \cmark & \cmark & \cmark & \cmark & \cmark & 1/GPL/S(3), F(4), W(-), C(6)/2023                                                               \\
		\ECSNeT \cite{ecsnet}                              & DES                                                                    & \CPP               & \cmark & \cmark & \cmark & \cmark & \cmark & \cmark & \cmark & \cmark & 17/None/S(9), F(7), W(3), C(1)/2020                                                             \\
		\SimGrid \cite{simgrid}                            & DES                                                                    & \CPP \Java \Python & \cmark & \cmark & \cmark & \cmark & \cmark & \cmark & \cmark & \cmark & 1288/LGPL/S(162), F(90), W(17), C(76)/2001\footnotemark \\
		\StepONE \cite{stepone}                            & DES                                                                    & \Java              & \cmark & \cmark & \xmark & \cmark & \cmark & \cmark & \cmark & \cmark & 13/GPL/S(3), F(0), W(4), C(1)/2020                                                              \\
		\hline
		\PeerSim \cite{peersim}                            & DES                                                                    & \Java              & \xmark & \xmark & \cmark & \cmark & \cmark & \xmark & \xmark & \xmark & 826/LGPLv2/in SourceForge\footnotemark/2009              \\
		\CloudSim (Plus) \cite{cloudsim, cloudsimplus}     & DES                                                                    & \Java              & \cmark & \xmark & \cmark & \cmark & \xmark & \xmark & \cmark & \xmark & 6.1k/Apache/S(727), F(474), W(66), C(6)/2011                                                    \\
		\CloudSimExpress \cite{cloudsimexpress}            & DES                                                                    & \Java \YAML        & \cmark & \xmark & \cmark & \cmark & \xmark & \cmark & \cmark & \xmark & 0/GPL/S(3), F(1), W(11), C(1)/2023                                                              \\
		\CloudSimPlusAutomation \cite{cloudsim-automation} & DES                                                                    & \Java \YAML        & \cmark & \xmark & \cmark & \cmark & \xmark & \cmark & \cmark & \xmark & 2/GPL/S(27), F(19), W(3), C(2)/2014                                                             \\
		\CloudSimSDN \cite{cloudsimsdn}                    & DES                                                                    & \Java              & \cmark & \cmark & \cmark & \cmark & \xmark & \xmark & \cmark & \xmark & 40/GPL/S(86), F(59), W(9), C(3)/2019                                                            \\
		\CloudSimPy \cite{cloudsimpy}                      & DES                                                                    & \Python            & \xmark & \xmark & \cmark & \xmark & \xmark & \xmark & \cmark & \xmark & 4/MIT/S(197), F(70), W(2), C(2)/2019                                                            \\
		\EdgeCloudSim \cite{edgecloudsim}                  & DES                                                                    & \Java              & \cmark & \cmark & \cmark & \cmark & \cmark & \xmark & \cmark & \xmark & 565/GPL/S(405), F(225), W(32), C(2)/2018                                                        \\
		\faassim \cite{faas}                               & DES                                                                    & \Python            & \cmark & \cmark & \cmark & \cmark & \cmark & \cmark & \cmark & \xmark & 4/MIT/S(58), F(16), W(3), C(5)/2023                                                             \\
		\iFogSim \cite{ifogsim2}                           & DES                                                                    & \Java              & \cmark & \cmark & \cmark & \cmark & \xmark & \xmark & \cmark & \xmark & 1.8k/None/S(123), F(72), W(17), C(2)/2022                                                       \\
		\IoTSimEdge \cite{iotsim-edge}                     & DES                                                                    & \Java              & \cmark & \cmark & \cmark & \cmark & \cmark & \xmark & \cmark & \cmark & 127/None/S(24), F(8), W(8), C(2)/2020                                                           \\
		\IoTSimOsmosis \cite{osmosis}                      & DES                                                                    & \Java              & \cmark & \cmark & \cmark & \cmark & \cmark & \xmark & \cmark & \cmark & 48/LGPL/S(8), F(7), W(1), C(1)/2021                                                             \\
		\LEAF \cite{leaf}                                  & DES                                                                    & \Java \Python      & \cmark & \cmark & \cmark & \cmark & \xmark & \xmark & \cmark & \xmark & 20/MIT/S(97), F(12), W(13), C(2)/2021                                                           \\
		\MobFogSim \cite{mobfogsim}                        & DES                                                                    & \Java              & \cmark & \cmark & \cmark & \cmark & \xmark & \xmark & \cmark & \xmark & 130/GPL/S(16), F(16), W(5), C(1)/2020                                                           \\
		\PureEdgeSim \cite{pureedgesim}                    & DSE                                                                    & \Java              & \cmark & \xmark & \cmark & \cmark & \cmark & \cmark & \cmark & \cmark & 47/GPL/S(166), F(72), W(4), C(4)/2019                                                           \\
		\RECAPDES \cite{recapdes}                          & DES\footnotemark                                                       & \Java              & \cmark & \xmark & \cmark & \cmark & \cmark & \xmark & \cmark & \xmark & 4/GPL/in Bitbucket\footnotemark/2020         \\
		\SatEdgeSim \cite{satedgesim}                      & DES                                                                    & \Java              & \cmark & \xmark & \cmark & \cmark & \cmark & \cmark & \cmark & \cmark & 17/EPL/S(37), F(10), W(1), C(2)/2020                                                            \\
		\SimFaaS \cite{simfaas}                            & DES                                                                    & \Python            & \cmark & \xmark & \xmark & \xmark & \xmark & \xmark & \cmark & \cmark & 26/MIT/S(20), F(10), W(3), C(1)/2021                                                            \\
		\SimEdgeIntel \cite{simedgeintel}                  & DES                                                                    & \Java \Python      & \cmark & \cmark & \cmark & \cmark & \cmark & \xmark & \cmark & \cmark & 23/MIT/S(15),F(7), W(3), C(1)/2021                                                              \\
		\VirtFogSim \cite{virfogsim}                       & DES                                                                    & \Matlab            & \cmark & \xmark & \cmark & \cmark & \xmark & \xmark & \cmark & \cmark & 20/MIT/S(4), F(4), W(1), C(1)/2019                                                              \\
		\YAFS \cite{yafs}                                  & DES                                                                    & \Python            & \cmark & \xmark & \cmark & \cmark & \cmark & \cmark & \cmark & \cmark & 219/MIT/S(97), F(69), W(12), C(7)/2019                                                          \\
		\hline
		\MaxiNet \cite{maxinet}                            & Emu                                                                    & \Python            & \xmark & \xmark & \xmark & \cmark & \cmark & \xmark & \cmark & \cmark & 233/Custom/S(88), F(41), W(17), C(8)/2014                                                       \\
		\Cooja \cite{cooja}                                & Emu                                                                    & \Java \CPP         & \cmark & \cmark & \cmark & \cmark & \cmark & \cmark & \cmark & \cmark & 292/BSD/S(28), F(46), W(9), C(18)/2009                                                          \\
		\MinNet \cite{mininet}                             & Emu                                                                    & \Python \Shell     & \cmark & \cmark & \cmark & \cmark & \cmark & \xmark & \cmark & \cmark & 2.7k/BSD/S(5.6k+), F(1.9k+), W(312), C(69)/2010                                                 \\
		\MiniNDN \cite{minindn}                            & Emu                                                                    & \Python \Shell     & \cmark & \cmark & \cmark & \cmark & \cmark & \xmark & \cmark & \xmark & np\footnotemark/GPL/S(66), F(39), W(9), C(22)/np                                                \\
		\Phantom \cite{shadow2}                            & DES                                                                    & \Rust \CPP         & \cmark & \cmark & \cmark & \cmark & \cmark & \xmark & \cmark & \xmark & 225/Custom/S(1.4k), F(235), W(50), C(42)/2011                                                   \\
		\StarryNet \cite{starrynet}                        & Emu                                                                    & \Python            & \cmark & \cmark & \cmark & \cmark & \cmark & \xmark & \cmark & \xmark & 16/MIT/S(57), F(5), W(2), C(3)/2023                                                             \\
		\CORE \cite{core}                                  & Emu                                                                    & \Python            & \xmark & \xmark & \xmark & \xmark & \xmark & \xmark & \xmark & \cmark & 429/BSD/S(610), F(154), W(37), C(28)/2009                                                       \\
		\DFaaS \cite{dfaas}                                & Emu                                                                    & \Python            & \cmark & \cmark & \cmark & \cmark & \cmark & \xmark & \cmark & \xmark & 22/None/S(16), F(2), W(3), C(7)/2021                                                            \\
		\EmuFog \cite{emufog}                              & Emu                                                                    & \Kotlin            & \xmark & \xmark & \xmark & \xmark & \cmark & \cmark & \xmark & \xmark & 155/MIT/S(23), F(9), W(6), C(2)/2017                                                            \\
		\Fogify \cite{fogify}                              & Emu                                                                    & \Python \YAML      & \cmark & \cmark & \cmark & \cmark & \cmark & \cmark & \cmark & \xmark & 29/Apache/S(24), F(8), W(8), C(2)/2020                                                          \\
		\hline
	\end{tabular}

	\begin{minipage}{18cm}
		\smallskip
		\textsuperscript{3} 
		DES: Discrete-Event Simulator, Emu: Emulator, Hybrid: Support both simulation and emulation.\\
		\textsuperscript{4} 
		Java: \Java, C/C++: \CPP, Python: \Python, YAML: \YAML, Kotlin: \Kotlin, Rust: \Rust, Matlab: \Matlab, Shell: \Shell. \\
		\textsuperscript{5} 
		SEV: Service (or application) layer,
		FAC: Facility layer between service and communication,
		HW: Hardware (can implement different hardwares, such as networking and computing nodes),
		NET: Network,
		CONF: Configuration file,
		SCEN: Scenario script,
		LOG: Logging Component,
		VIS: Visualization.\\
		\textsuperscript{6}citations/licenses/project information (\underline{S}tars, \underline{F}orks, \underline{W}atchers, \underline{C}ontributors)/published or released year. \\
	    \textsuperscript{7}statistics consists of GitHub and FramaGit.\\
        \textsuperscript{8} https://sourceforge.net/projects/peersim.\\
        \textsuperscript{9} RECAP also has a discrete time simulator \cite{recapdes}.\\
        \textsuperscript{10} https://bitbucket.org/RECAP-DES/recap-des/src/master.\\
        \textsuperscript{11} No referenced paper.
  
	\end{minipage}
\end{table*}

\section{Future Directions}
\label{sect_future_directions}
We provide four tables (\tablename~\ref{tab:compt-paradigms}-\tablename~\ref{tbl_usability}) as a quick reference for readers to find suitable tools for verifying their ideas. Based on our analysis, we illustrate some future directions in this section.

\paragraph{Computing Paradigms Integration}
As demonstrated, recent simulators and emulators support various computing paradigms, ranging from IoT, Edge, Fog, to Cloud. These tools are developed either from networking (packet-level) or computing (application-level) perspectives, with a trend towards integration, as seen in \CloudSimSDN and \EasiEI. Establishing a unified classification among these tools is challenging. Future research should focus on identifying and structuring the unique features of different simulators and emulators to facilitate their comparison and highlight their strengths. Meanwhile, providing common libraries such as scheduling to reduce code redundancy are also needed.

\paragraph{Packet Processing in Application-level Simulators}
Recent research optimizes performance by jointly considering both data contents and task (or application) computing optimization, such as the age of information (AoI) in edge computing, where data freshness impacts task processing effectiveness \cite{yates2021age, chiariotti2022query, RN94}, computing while transmitting \cite{RN1315}, and compute first networking \cite{cfn, 9862944}. Although various scheduling methods are analyzed in application-level simulators, the feature of packet-level processing is often overlooked. Future work should integrate packet-level processing features into application-level simulators to enhance their capability.

\paragraph{Edge Environments Simulation Support}
Edge environments, such as weather conditions, act as intermediate factors that can influence computing performance. For instance, in ITS simulators, factors like fog can negatively impact vehicle sight \cite{artery}. Similarly, temperature and humidity can affect computing processes and capabilities. Therefore, future simulators should incorporate edge environment simulation to provide a more comprehensive analysis of these factors.

\paragraph{User-defined Metric Interface Support}
Proposing a new algorithm or method often involves introducing new indicators or metrics, such as various reliability measures, from application-level time thresholds to packet-level data transmission errors. Providing well-defined interfaces for user-defined metrics could be beneficial. Additionally, utilizing database tools, such as SQLite in \NSthree, can facilitate agile output analysis and enhance usability.

\paragraph{Scenario Scripts and Visualization Support}
The diversity of edge devices in \cnc necessitates diverse scenarios. Providing a unified way to process several sub-applications or tasks is feasible and beneficial. For example, \StepONE can help users manage the complex orchestration of multiple services, simplifying the process and improving usability. Future developments should focus on enhancing scenario scripting and visualization support to aid in the effective simulation and analysis of various edge computing scenarios.

\section{Conclusion}
\label{sect_conclusion}
This survey highlights the diverse range of open-source edge computing simulators and emulators available for researchers. These tools support various computing paradigms, from traditional networking to emerging concepts like fog and serverless computing. 
By providing a detailed comparison and analysis, this paper aims to assist researchers in selecting the most suitable tools for their specific requirements, thereby facilitating the validation and development of advanced computing and networking technologies. Future research should focus on further integrating computing paradigms, improving packet processing in application-level simulators, and enhancing support for edge environments and user-defined metrics.

\balance
\bibliography{reference}
\bibliographystyle{plain}

\end{document}